\documentclass{article}

\usepackage{arxiv}

\usepackage[utf8]{inputenc} 
\usepackage[T1]{fontenc}    
\usepackage{hyperref}       
\usepackage{url}            
\usepackage{booktabs}       
\usepackage{amsfonts}       
\usepackage{nicefrac}       
\usepackage{microtype}      
\usepackage{lipsum}
\usepackage{graphicx}

\title{Text analysis in financial disclosures}

\author{
    Sridhar Ravula
   \\
    Department of Analytics \\
    Harrisburg University of Science and Technology \\
  Harrisburg, PA 17101 \\
  \texttt{\href{mailto:sravula@my.harrisburgu.edu}{\nolinkurl{sravula@my.harrisburgu.edu}}} \\
  }

\newlength{\csllabelwidth}
\newlength{\cslhangindent}
\newenvironment{cslreferences}%
  {\setlength{\parindent}{0pt}%
  \everypar{\setlength{\hangindent}{\cslhangindent}}\ignorespaces}%
  {\par}
 {
  \setlength{\parindent}{0pt}
  \ifodd #1 \everypar{\setlength{\hangindent}{\cslhangindent}}\ignorespaces\fi
  \ifnum #2 > 0
  \setlength{\parskip}{#3\baselineskip}
  \fi
 }%
 {}
\usepackage{calc} 

\begin{document}
\maketitle

\def\tightlist{}

\begin{abstract}
Financial disclosure analysis and Knowledge extraction is an important
financial analysis problem. Prevailing methods depend predominantly on
quantitative ratios and techniques, which suffer from limitations like
window dressing and past focus. Most of the information in a firm's
financial disclosures is in unstructured text and contains valuable
information about its health. Humans and machines fail to analyze it
satisfactorily due to the enormous volume and unstructured nature,
respectively. Researchers have started analyzing text content in
disclosures recently. This paper covers the previous work in
unstructured data analysis in Finance and Accounting. It also explores
the state of art methods in computational linguistics and reviews the
current methodologies in Natural Language Processing (NLP).
Specifically, it focuses on research related to text source, linguistic
attributes, firm attributes, and mathematical models employed in the
text analysis approach. This work contributes to disclosure analysis
methods by highlighting the limitations of the current focus on
sentiment metrics and highlighting broader future research areas.
\end{abstract}


\hypertarget{introduction-and-background}{%
\section{Introduction and
background}\label{introduction-and-background}}

The steady flow of accurate and requisite information is essential for
market efficiency. The result of this information flow is a far more
active, efficient, and transparent capital market that facilitates
capital formation so important to the nation's economy. The October 1929
stock market crash plummeted public confidence in the markets. Investors
and the banks who had loaned to them lost money in the following Great
Depression. Based on the problems identified and proposed solutions,
Congress passed the Securities Act of 1933 and the Securities Exchange
Act of 1934. Through these acts, Congress created the SEC to restore
investor confidence in capital markets by providing more reliable
information and transparent rules.

To enhance and facilitate market information flow, the SEC offers the
public a wealth of information, including the EDGAR database of
disclosure documents that public companies file with the
Commission.\footnote{SEC History
  \url{https://www.sec.gov/about/reports/sec-fy2015-agency-mission-information.pdf}}
This information empowers users to focus on the primary goal, i.e.,
financial modeling and security analysis, rather than on the information
search and acquisition process.

Investors and analysts place great emphasis on security analysis and
valuation because of the potential excess returns on capital and the
downside risks. Research in this domain is potentially valuable because
market inefficiencies can result in volatility and crashes, costing the
economy billions of dollars. Analysts extensively use public firm's
disclosures as a source of information.

\hypertarget{accounting-metrics-models}{%
\subsection{Accounting metrics models}\label{accounting-metrics-models}}

Investors and analysts traditionally depended on quantitative
information like accounting metrics for decision making. Multiple
attributes of these accounting metrics drove this trend. FACC and
accounting standards laid out what variables to be measured and
disclosed. Gathering, processing, and analyzing these quantitative
metrics was easy. Many free and commercial data providers automated data
gathering, and publish these metrics.

However, these metrics do not always reveal the firm's current status
and are not a good indicator of the future. They suffer from
shortcomings like window dressing and retrospective focus

\hypertarget{window-dressing}{%
\paragraph{Window dressing}\label{window-dressing}}

Managers have the motivation and opportunity to report ``desired''
numbers than actuals. Managers know the financial ratios monitored by
markets and regulators. Financial ratios can be fudged with temporary
transactions, improving them at the time of reporting. Managers can use
accounting discretion to value assets or to report inaccurate
information. Accounting metrics analysis ignores the possibility of
``contaminated source.'' As the purpose of the research is to identify
material changes, and firms with negative material changes have more
incentives to window dress, inaccurate disclosures is a significant
concern. Sufficient evidence exists for window dressing through
commissions and omissions. Rajan, Seru, and Vig (2015) showed that banks
did not report information regarding the deteriorating quality of
borrowers' disclosures in the run-up to the subprime crisis. Huizinga
and Laeven (2012) said that banks overstated the value of their
distressed real estate assets and regulatory capital.

\hypertarget{retrospective-focus}{%
\paragraph{Retrospective focus}\label{retrospective-focus}}

Accounting metrics analysis focuses on past performance. Investment and
lending activities are forward-looking. Accounting information captures
reality with lags. Some of the measures might be outdated by the time
they are used in the analysis. Future forecasts based on these metrics
also suffer from the same disadvantages.

\hypertarget{missing-variables}{%
\paragraph{Missing variables}\label{missing-variables}}

Firms are complex social organizations, and numerous social, political,
and economic factors influence their performance. A firm's performance,
survival, and profitability are affected by multiple factors. Companies
do not measure all of these factors. Also, unless regulatorily mandated,
firms might not report them. Regulators keep adding new reporting
requirements to overcome this, but often fighting the last war and
solving the past crisis. As a result, investors do not have access to
some of the essential variables.

Window dressing, retrospective focus, and missing variables impact
models based on accounting metrics. Regulators and investors who rely on
such models have been impacted adversely in the past due to model
failures (Rajan, Seru, and Vig (2015)). Hence researchers started paying
more attention to alternative approached like market based models and
textual analysis of disclosures.

\hypertarget{market-based-models}{%
\subsection{Market based models}\label{market-based-models}}

Another approach for bankruptcy prediction is using market-based
information. Classical efficient market theory and later option pricing
theories assume that all available information is reflected in market
prices. Under those conditions, accounting-based metrics do not have
additional information over and above market prices. More specifically,
a suitable market-based measure will reflect all available information
about bankruptcy probability. Hillegeist et al. (2004) developed a
prediction model based on market information, using option pricing
theory derived implied volatility. This model outperformed the Altman
(1968) z score model. Subsequently, numerous attempts have been made to
replicate these results. Wu, Gaunt, and Gray (2010) provides a
comparison of accounting and market-based models, along with others.
They conclude that the Hillegeist et al. (2004) model performs better
than the Z score model but is inferior to models that include
non-traditional metrics. Similarly, Tinoco and Wilson (2013) concluded
that accounting metrics based models and market-based models are
complimentary.

\hypertarget{text-analysis}{%
\subsection{Text analysis}\label{text-analysis}}

Disclosure documents are rich with text information and provide a better
insight into the future of the firm. Loughran and Mcdonald (2016) has
demonstrated the relationship between disclosure text features and firm
attributes.

However, unstructured text analysis remains difficult and still requires
juggling many manual language processing tasks and technologies. Text
information in disclosures prevents the computerized automation of
security analysis. Engelberg (2008) noted that while text-based soft
information has predictive power, this information comes at a higher
cost due to processing challenges. As the information volume is high,
analyst teams cannot promptly cope with it, let alone individual
investors. Investors need the ability to extract reliable information
and knowledge from disclosure texts with minimum manual effort.

Computational text analysis and Natural Language Processing (NLP)
methods can help in extracting this information. Text analysis in
finance can provide insights into the managerial motivation to disclose
or obfuscate information, the circumstances for obfuscation, and
impacted linguistic features. Natural Language Processing offers
insights into the language models, information extraction, topic
modeling, and new methodologies for measuring information content.

Researchers have started analyzing text content in disclosures recently.
This survey covers the previous work in unstructured data analysis in
Finance and Accounting. It also explores the state of art methods in
computational linguistics and reviews the current methodologies in
Natural Language Processing (NLP).

It consists of four sub-chapters as below.

\begin{enumerate}
\def\labelenumi{\arabic{enumi}.}
\tightlist
\item
  EDGAR is the primary information source.
\item
  Text analysis in Finance.
\item
  Firm attributes.
\item
  Modeling approaches used.
\end{enumerate}

\hypertarget{edgar-as-the-primary-source-of-financial-disclosure-information}{%
\section{EDGAR as the primary source of financial disclosure
information}\label{edgar-as-the-primary-source-of-financial-disclosure-information}}

SEC regulations and oversight guide disclosure practices to increase
transparency and reduce the likelihood of individual stock price
crashes. To enhance and facilitate market information flow, the SEC
offers the public a wealth of information, including the disclosure
documents that public companies must file with the Commission.

While several information sources are relevant for the Financial domain,
the Electronic Data Gathering, Analysis, and Retrieval system (EDGAR)
filings represent the first-source database for investors doing
fundamental research on security valuation. SEC introduced EDGAR in the
early 90s and gradually made it mandatory for most listed companies to
file their disclosures through EDGAR. Investors rely on EDGAR
disclosures, and these disclosures are effective in improving market
efficiency. Though firms file hundreds of types of disclosure forms,
annual and quarterly filings (10-K, 10-Q), as well as material change
filings (8-K), have the bulk of user attention.

If analysts can access EDGAR filings, they can incorporate the
information into forecasts and security prices. While analysts have
historically used financial disclosures, EDGAR's advent made it possible
for individual investors to do the same. Christensen, Heninger, and
Stice (2013) found that revisions in analysts' one-quarter-ahead
earnings forecasts around SEC filings dates in both the pre-EDGAR and
EDGAR periods were significant. However, the stock price reactions to
SEC filings are meaningful only in the EDGAR period, indicating that
individual investors can incorporate the information in their price
expectation.

\hypertarget{edgar-filings--market-reaction}{%
\subsection{EDGAR filings- market
reaction}\label{edgar-filings--market-reaction}}

Numerous other studies researched and established market reactions to
SEC EDGAR filings. Stice (1991) studied the impact of early filings.
Bühner and Möller (1985) assessed the information content of corporate
disclosures relating to decisions to adopt a multi-divisional structure.
Asthana and Balsam (2001) examined the effect of filing form 10-K on
EDGAR when compared to the traditional method of filing. Yang (2015)
studied the disclosure and valuation of foreign cash holdings. Li et al.
(2009) found significant market reaction during quarterly reports if the
filing coincides with the first public disclosure of earnings, although
earnings releases do not subsume that for 10-K reports. You and Zhang
(2011a) concluded that among large firms, investors under-react more to
the information contained in 10-K filings than earnings announcements,
and underreaction to earnings announcements tends to be stronger for
small firms than large firms. You and Zhang (2011b) observed that
investors use information from earnings reports and 10-K disclosures
differently.

\hypertarget{delay-in-filings}{%
\subsubsection{Delay in filings}\label{delay-in-filings}}

Firms' inability to file disclosure on time has information content.
Duarte‐Silva et al. (2013) studied the market reaction to earnings delay
announcements and concluded that these delays provide a signal of
financial performance deterioration. Khalil et al. (2017) observed
similar effects in bond markets as bond markets react negatively to late
filing announcements of 10-Ks and 10-Qs. However, the cost of
non-compliance with SEC filing requirements vary, based on whether the
delay is due to a systems-related issue or not. Investors' ability to
discern the reasons behind delays results in differential market
response. Cao et al. (2010) investigated the reasons stated for delayed
filings on Form 12b-25 and found evidence that negative market returns
are associated with filing delays caused by Information System issues,
SOX implementations, and SEC investigations.

\hypertarget{non-periodic-disclosures}{%
\subsection{Non-periodic disclosures}\label{non-periodic-disclosures}}

Over time, the SEC expanded the items that firms have to report and
accelerated these reports' timelines. Form 8-K, which became effective
on August 23, 2004, is the ``current report'' companies must file to
announce major events that shareholders needed to know. As a result,
investors need not wait for periodic reports for time-sensitive
information. However, Lerman and Livnat (2010) observed that this has
not resulted in a reduction of information content in periodic reports.
Market reactions to 8-Ks indicate that investors may still rely more on
10-K and 10-Q reports to interpret material events' effects.

Firms' inability to file Form 8-K on time may indicate internal control
weakness. In a study covering a sample of 118,863 8-K filing event
observations, Holder et al. (2016) found a negative relation between the
likelihood of a material internal control weakness and the timeliness
and compliance of 8-K filings. The market reacts to SEC 8-K filings on
and between the event's date and the date of filing. Ben-Rephael et al.
(2017) showed significant abnormal attention paid by institutional
investors on both the filing date and the event date. The same is not
the case with retail investors. Considering the high volume of 8-K
disclosures and relatively less information content than periodic
filings, retail investors might not be paying sufficient attention to
the 8-K filing. Most retail investors use manual methods to extract
information from EDGAR, and large volume is a hindrance. As most price
discovery occurs during the pre-filing period when institutional
investors pay attention, retail investors might be disadvantaged. The
automatic processing of such filings for information extraction can
benefit retail investors.

\begin{table}

\caption[Top 20 EDGAR Filings]{\label{tab:unnamed-chunk-2}Top 20 EDGAR filings till Dec 2018}
\centering
\begin{tabular}[t]{lll}
\toprule
Description & Form.Name & Grand.Total\\
\midrule
Changes in ownership & 4 & 6,757,053\\
Current report filing & 8-K & 1,549,474\\
5\% passive ownership triggers amendments & SC 13G/A & 655,696\\
Initial Ownership Report & 3 & 613,615\\
Quarterly report & 10-Q & 574,774\\
\addlinespace
Definite materials & 497 & 417,177\\
Curret report of Foreign issuer & 6-K & 379,844\\
5\% passive ownership triggers & SC 13G & 369,518\\
Change on prospectus & 424B3 & 276,705\\
Quarterly hodings, institutional managers & 13F-HR & 238,794\\
\addlinespace
Prospectus filed pursuant to Rule 424(b)(2) & 424B2 & 236,485\\
5\% active ownership triggers amendments & SC 13D/A & 222,739\\
Changes in ownership amendments & 4/A & 217,095\\
Offering made without registration & D & 211,275\\
Annual report on ownership changes & 5 & 199,887\\
\addlinespace
A correspondence & CORRESP & 186,948\\
Upload Notification & UPLOAD & 183,620\\
Annual report & 10-K & 183,148\\
Post-effective amendment & 485BPOS & 171,995\\
Summary Prospectus & 497K & 169,519\\
\bottomrule
\end{tabular}
\end{table}

\hypertarget{edgar-usage}{%
\subsection{Edgar usage}\label{edgar-usage}}

While EDGAR is useful, there were concerns by stakeholders about the
cost-benefit balance. To better understand the usage of EDGAR company
filings, SEC started releasing EDGAR Log File Data Set. The Division of
Economic and Risk Analysis (DERA) has assembled, and published
information on all internet search traffic for EDGAR filings through
SEC.gov, and current data covers the period February 14, 2003, through
June 30, 2017. Researchers have studied the timing and investors'
accessing patterns of financial filings. Investors request millions of
filings from EDGAR each week, showing a high preference for 10-K, 10-Q,
and 8-K, along with insider trading disclosures filed on Form 4.

The timing of investor requests indicates that investors commonly
request historical disclosures filed in prior periods. This demand is
high when the stock is under-performing, i.e., past and current abnormal
stock returns are lower. Drake, Roulstone, and Thornock (2012) found
that investors access mandatory financial filings during news release
periods, and the demand increases during times of negative news and
increased uncertainty about the firm's business.

Jackson and Mitts (2014) demonstrated that investors' ability to access
and process market-moving information before others results in trading
profits. Based on EDGAR server access traffic from 2008-2011, Drake,
Roulstone, and Thornock (2015) found that information acquisition via
EDGAR positively influences market efficiency. Ryans (2017) analyzed the
same data set from a different perspective and demonstrated the
differences between human requests and financial robots.

Other research findings on EDGAR access traffic

\begin{itemize}
\tightlist
\item
  Average user accesses the database infrequently.
\item
  Users access specific periodic filing types such as 10-K and 10-Q.
\item
  Few users access EDGAR almost daily and access numerous filings.
\item
  EDGAR activity is positively related with

  \begin{itemize}
  \tightlist
  \item
    corporate events

    \begin{itemize}
    \tightlist
    \item
      restatements
    \item
      earnings announcements
    \item
      acquisition announcements
    \end{itemize}
  \item
    poor stock performance
  \item
    EDGAR activity is distinct from other investor interest proxies and
    has unique information content.
  \end{itemize}
\end{itemize}

A 10-K filing is a very long, complicated document, and investors need
to spend hours comprehending the same. As a result, individual investors
might be relying on analysts and secondary research products for the
fundamental analysis. Loughran and McDonald (2017) found that the
average publicly traded firm has its annual report requested only 28.4
total times immediately after the 10 K-filing. They concluded that
investors generally are not doing fundamental research on stocks. The
low number of access requests indicate the challenges of processing
EDGAR filings and suggest the requirement for automatic information
extraction and knowledge discovery.

By linking EDGAR server activity to analysts' brokerage houses, Gibbons,
Iliev, and Kalodimos (2018) concluded that analysts rely on EDGAR in
26\{\%\} of their estimate updates. They found that fundamental research
is associated with a significant reduction in analysts' forecasting
error relative to their peers.

These studies indicate that investors and analysts access mandatory
disclosures through EDGAR and the information discovery impacts price
formation. They demonstrate the importance of fast processing of
information as a delay in the order of seconds has a significant
opportunity cost. Also, they show the difficulties in extracting data
from EDGAR in real time, especially for retail investors. Considering
that historical filings are frequently accessed, and firms file 2
million disclosures every year, automated information extraction is a
must for timely dissemination and price discovery.

\hypertarget{xbrl}{%
\subsection{XBRL}\label{xbrl}}

To facilitate automated information extraction, SEC has adopted the
eXtensible Business Reporting Language (XBRL) and enhanced the EDGAR
database. SEC initiated XBRL in 2009 and mandated that all firms use
XBRL by 2011. Later, it extended the due date to 2014. In XBRL, filers
tag their financial statements with elements from a taxonomy that
defines the reporting concepts so that the information consumers can
understand XBRL files (Debreceny et al. (2011)). According to
Henselmann, Ditter, and Scherr (2013), this enables the gathering of
accounting numbers to be fully automatic in a database-like manner,
which provides vast opportunities for financial analysis.

Numerous researchers studied the impact of XBRL across various aspects
of the financial information environment, like market efficiency, price
discovery, and volatility. Kim, Lim, and No (2012) findings show an
increase in information efficiency, a decrease in event return
volatility, and a reduction of change in stock returns volatility for
428 firms (1,536 10-K and 10-Q filings) post-XBRL disclosure. Their
study also showed that XBRL mitigates information risk in the market,
especially during increased uncertainty in the information environment.
Mangold et al. (2013) analyzed a sample of 671 amended filings for XBRL
from 2005 to 2011 and found a significant market reaction to the XBRL

After initial years of XBRL implementation, there have been complaints
by some Fortune 500 companies that XBRL filings have not proven useful
and have advocated for the SEC to scale down its requirements. However,
research continued to show the positive impact of XBRL. Analysts
under-react to initial information releases if they expect subsequent
follow-up disclosures to have better quality information. Dontoh and
Trabelsi (2015) found that the market reacted significantly less to
earnings announcements of firms that issued XBRL filings than a matched
sample that did not give filings. They also found that firms that issued
XBRL filings exhibited significantly lower excess return volatility than
non-filers and concluded that overall, XBRL has been useful in improving
market efficiency.

Some researchers tried to develop tools to leverage XBRL information for
identifying new information in the market. Henselmann, Ditter, and
Scherr (2013) proposed abnormal digit distributions at the firm-year
level to identify firms indulging in earnings management. Hoitash and
Hoitash (2018) suggested a ``count of accounting items (XBRL tags)'' in
10-K filings as a measure of accounting reporting complexity (ARC). This
complexity aspect can increase the likelihood of mistakes, incorrect
GAAP application and ultimately lead to less credible financial reports.

\hypertarget{shortcomings-of-xbrl}{%
\subsubsection{Shortcomings of XBRL}\label{shortcomings-of-xbrl}}

While the above studies indicated the benefits of XBRL, researchers
studied the complexity introduced by XBRL vs.~the corresponding
benefits. When SEC introduced XBRL, it believed that this new
search-facilitating technology would reduce informational barriers and
asymmetry that separate smaller, less-sophisticated investors from
larger, more sophisticated investors. If larger investors gain
significant benefits from XBRL through their superior resources and
abilities, smaller investors will be disadvantaged, and information
asymmetry is likely to increase. Blankespoor, Miller, and White (2014)
studied this question and observed higher bid-ask spreads for XBRL
adopting firms. While XBRL may have reduced investors' data aggregation
costs, it may not have leveled the initial years' informational playing
field.

The XBRL mandate intends to streamline the financial reporting pipeline
by providing a standard dictionary for collecting, collating, and
analyzing financial information on the Web. However, the current lack of
suitable XBRL interoperability prevents the realization of the mandate's
potential. SEC provided GAAP taxonomy for filing the reports in XBRL
format. The U.S. GAAP taxonomy's design supports standard reporting
practices and U.S. GAAP disclosure requirements. It also allowed the
filer creates an extension element if taxonomy elements for each
disclosure concept are not present. While appropriate extensions can
provide decision-relevant information, unnecessary extensions, when
suitable elements exist in foundation taxonomy, create redundancy and no
information content. Debreceny et al. (2011) analyzed extensions made in
a subset of XBRL filings between April 2009 and June 2010. It concluded
that forty percent of these extensions were unnecessary, as semantically
equivalent elements were already in the U.S. GAAP taxonomy. New
concepts, many of which were variants of existing features, accounted
for 30 percent of the extensions.

Value addition from XBRL-tagged data will materialize only if the XBRL
statements are accurate and reliable. Bartley, Chen, and Taylor (2011)
studied errors in XBRL filings and evaluated the accuracy of early
voluntary filings. While improvements in the XBRL standard and related
technology mitigate specific errors, other errors related to
inexperience will persist.

The above results suggest that while XBRL has helped in information
discovery, it has not yet delivered the intended benefits. Streamlining
the financial reporting pipeline has remained a pipe dream. Also,
considering that significantly more information is present in the form
of text in disclosures, analysts cannot access it using traditional
structured data analysis. Hence there is a need for unstructured data
analysis.

This sub-chapter has demonstrated how EDGAR acts as a primary source of
financial disclosures and how the SECs attempts to enable easy
information extraction still suffer from shortcomings. The next
sub-chapter will analyze how analysts have attempted to extract
information from text using text analysis.

\hypertarget{text-analysis-in-finance}{%
\section{Text analysis in finance}\label{text-analysis-in-finance}}

This sub-section covers the previous work in unstructured data analysis
in Finance and Accounting. Most of the literature of these fields talks
in terms of sentiment. While researchers used dictionary-based methods,
recent success has been more due to probabilistic techniques. The
following literature review helps us identify which of the many
statistically motivated options are best for this problem. Finally, this
section examines several practical applications of the theoretical
techniques relevant to current work. Analysts and individuals pay
attention to text components of a firm's disclosures as company
disclosures traditionally have been accompanied by narrative disclosures
regarding the companies (Fisher et al. (2010)). Financial and accounting
researchers have been paying attention to text analysis research
problems, i.e., ``Efficient use of narrative textual documents.'' Prior
research used different information sources like corporate textual
disclosures in filings, accounting standards, and CEO statements for
various purposes, such as forecasting future performance, dictionary
development, and formalization (Loughran and Mcdonald (2016)).

As noted in the prior sections, financial reporting and disclosures'
objective is to ensure the availability of information about firms'
financial position to a wide range of users, including existing and
potential investors, financial institutions, employees, and the
government. Text analysis in Finance starts from the hypothesis that
there is a relation between linguistic properties of disclosures and
business performance (Smailović et al. (2018)). In recent years, text
analysis in Finance and accounting has seen a dramatic increase in
attention. Increased availability of technology for storing and
accessing data was the driver. This increased attention resulted in
better methods to process a large corpus of texts in real time in other
domains. Some of those methods have been attempted in Finance also.

The following sections contain a brief review of text analysis in
Finance and accounting, structured into three parts.

\begin{itemize}
\item
  Content analysis: This part covers ``sections and content from
  financial disclosures'' that researchers studied. These sections are
  the ``source of linguistic features'' that explain the business
  performance. There are more than 200 types of disclosure forms, and
  each has multiple sections. Understanding text content used in
  research helps us understand the researchers' objectives and the
  research gaps.
\item
  Linguistic Properties: Covers ``What features are extracted'' from
  financial disclosure text. While linguistic properties can explain
  business performance, suitable methods have to transform the text into
  linguistic measures.
\item
  Business performance attributes This part covers ``Which aspect of
  business performance'' the researchers tried to explain with
  linguistic variables.
\item
  Statistical methodology: This part covers what types of statistical
  tools and methods used in the research
\end{itemize}

\hypertarget{content-analysis}{%
\subsection{Content analysis}\label{content-analysis}}

Communication between corporate management and various interested
constituencies occurs continuously and in many forms. A traditional
statutory-based formal communication vehicle is the corporate annual
report. Although a less timely medium than other filings, the annual
report comprises a comprehensive database of past corporate
achievements, thereby facilitating the confirmation, revision, and
formation of readers' expectations about a company in which they have an
interest (Courtis (1998)) . The annual report will be more or less
useful depending on the extent to which, among other things, its content
is readable and understandable. One contemporary annual reporting trend
is for management to employ more narrative disclosures as part of the
overall communication package.

A typical financial disclosure has many sections. Each section has
relevance to one or many of the investor's research questions.
Extracting a relevant text segment is a computational challenge.
Information extraction deals with the problem of identifying the
required information and pulling from a source. The number of filings
with EDGAR has exceeded 18mn as of Dec 2018. Figure 1 shows the filings
trend over the year. With over a million new filings each year, manual
analysts have reached their capacity limits in exploring this dataset.
Systematic exploration of this dataset necessitates automation to
overcome overload.

\begin{figure}
\includegraphics[width=1\linewidth]{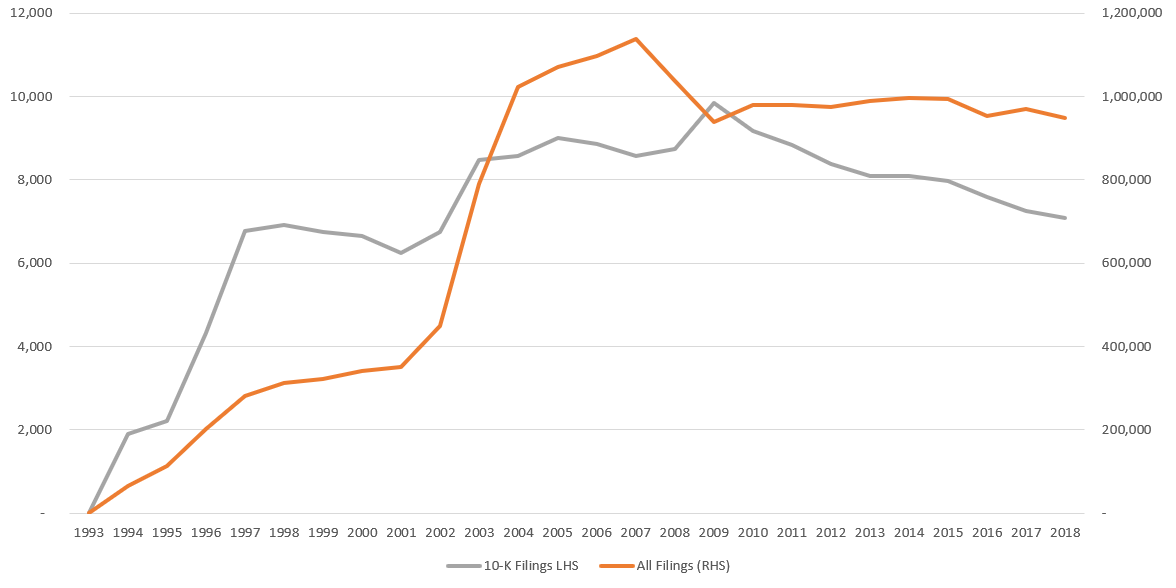} \caption{Number of annual EDGAR filings}\label{fig:Filings}
\end{figure}

Apart from the large volume of filings, some of the filings are lengthy,
further hampering analysts' ability to process and understand the
information. A significant portion of text in financial disclosures like
10-K reports is under managerial discretion and depends on how firms
respond to mandatory disclosure requirements. Other factors that
influence the length of filings include firms operating complexity and
disclosure redundancy. Cazier and Pfeiffer (2016) partitioned 10-K
length into the portions explained by these factors. While disclosure
redundancy and operational complexity explain roughly equal amounts of
variation in 10-K size, the remaining content explains a higher
proportion of variation supporting the managerial discretion theory. Due
to this, analysts need to give enough consideration in identifying using
sections of financial disclosures.

Researchers have approached this in different directions. Cong, Kogan,
and Vasarhelyi (2007) used the income statement section of 10-K filings
and a set of heuristic-based templates to train a system to process one
type of EDGAR filings in a single configuration. Another approach is to
construct domain-specific languages to provide modeling capabilities
tailored to a specific domain. Li and Zhao (2014) introduced a
domain-specific meta-modeling language with examples from different
fields, including financial disclosures. Matthies and Coners (2015)
evaluated two text analysis strategies - dictionary and statistical
approach and concluded that they complement each other. Han et al.
(2016) provides general data extraction and analysis resolution for
mining the business knowledge from EDGAR. Focusing on each company's
annual meeting date from the `DEF 14A' form, they automatically scanned
546,451 documents and extracted 82,872 annual meeting date records of
10,417 companies. Topic modeling for comparative text analytics is
another computational linguistic approach in banking. Chen et al. (2018)
compared and evaluated multiple topic modeling approaches in analyzing
U.S. banks' SEC filings by U.S. public banks and concluded that topic
modeling could be useful in financial decision making and risk
management.

Numerous research works have analyzed the complete content of
disclosures. Hendricks, Lang, and Merkley (2017) examined whether
textual attributes of firms' regulatory filings reflect CEO
characteristics. Buchholz et al. (2018) concluded that CEO's narcissism
explains the abnormal optimistic tone in financial disclosures.

\hypertarget{management-discussion-analysis-mda}{%
\subsubsection{Management Discussion \& Analysis
(MDA)}\label{management-discussion-analysis-mda}}

Management Discussion and Analysis (MDA) is a Part of a Firm's Overall
Disclosure Package. As MDA contains new and useful information about the
firm, investors use it for financial analysis purposes. Clarkson, Kao,
and Richardson (1999) presented evidence regarding the usefulness of
MD\&A. and on disclosure quality. Foster and Hussey (2006) trained a
proprietary algorithm to quantify companies' strategic orientation,
based on semantic patterns within the MDA section of 70,000 10-K
filings. Preparing MDA is time-consuming. Firms have an incentive to use
a template to reduce the effort while remaining compliant with
regulatory requirements. MDA information value diminishes if it does not
change significantly, even after material changes at the firm. SEC
expressed concern about in the late 2000s. Brown and Tucker (2011)
studied year over year changes in MDA and introduced a measure for
disclosure. They found that firms modify the MD\&A more in the presence
of more considerable economic changes, and the modification score is
positively related to stock price responses. After initial research work
on text-based models, other researchers evaluated if the text-based
analysis has excess explanatory power over quantitative models. There is
evidence that text-enhanced models are more accurate than models using
quantitative financial variables alone. Bochkay (2014) explored
methodologies for analyzing and incorporating text into quantitative
models. Using regularized regression methods, they examined whether
textual disclosures in the MDA help predict future earnings above and
beyond traditional financial factors. They found that Firms with broader
changes in future performance, adverse changes in future performance,
higher accruals, greater market capitalization, and lower Z-scores have
more informative MD\&As, suggesting that MD\&A content helps to reduce
uncertainty.

Numerous other researchers analyzed MDA to explain future stock
performance (Tao, Deokar, and Deshmukh (2018a)), future returns,
volatility, and firm profitability (Amel-Zadeh and Faasse (2016)),
bankruptcy (Yang, Dolar, and Mo (2018)), going-concern (Mayew,
Sethuraman, and Venkatachalam (2015),Enev (2017)), litigation risk
(Bourveau, Lou, and Wang (2018)), and incremental information over
earnings surprises, accruals and operating cash flows (OCF)(Feldman et
al. (2008),Feldman et al. (2010)).

Another part of MDA that attracted research interest was Forward-looking
statements (FLSs), especially from IPO prospectuses. FLSs provide
prospective information about the company's future growth and
performance. Tao, Deokar, and Deshmukh (2018a) evaluated the
relationship between features extracted from FLSs and IPO valuation.
They proposed an analytical pipeline for identifying FLSs and extracting
linguistics features, including topics, sentiments, readability,
semantic similarity, and general text features. They concluded that FLS
features are more predictive for pre-IPO as compared to post-IPO
valuation prediction. F. Li (2010b) found that firms with better current
performance, lower accruals, smaller size, lower market-to-book ratio,
less return volatility, lower MDA Fog index, and long history tend to
have more positive FLSs. Also, the average tone of the FLS is positively
associated with future earnings, even after controlling for other
determinants of future performance.

\hypertarget{financial-notes}{%
\subsubsection{Financial notes}\label{financial-notes}}

Another important section in filings is Financial notes. It is a general
practice for managements to explain large deviations, year on year
changes, and material treatment changes through financial notes.
Analysts use these notes to compute accounting adjustments to correct
financial statements. This mechanism helps in incorporating this
information into stock prices.

De Franco, Franco Wong, and Zhou (2011) examined financial statement
notes in 10-K filings and observed that stock returns are positively
related to accounting adjustments. They also found that equity analysts
are likely to update target price estimates in proportion to an increase
in the adjustments' magnitude. The revisions are consistent with the
sign and extent of the adjustments. Amel-Zadeh and Faasse (2016)
concluded that investors react in a timely and robust manner to textual
characteristics of the MDA to textual attributes of the financial notes.

The above studies focused on accessing and extracting text information
from EDGAR filings for further analysis. The following section covers
what linguistic features have been studied, i.e., sentiment, tone, and
readability.

\hypertarget{language-analysis}{%
\subsection{Language analysis}\label{language-analysis}}

This section surveys the independent variables used in the textual
analysis of financial disclosures.

\hypertarget{sentiment}{%
\subsubsection{Sentiment}\label{sentiment}}

Sentiment analysis of firm disclosures involves an automated process of
understanding management's opinion about various aspects of a firm's
economic and financial prospects. Management opinions provide
qualitative information on the entity's financial status apart from
quantitative data. Investors and analysts consider and incorporate this
information into financial decisions. Chen et al. (2011) built tagging
models based on the conditional random field (CRF) techniques to study
opinion patterns in U.S. financial statements. Others have studied the
impact of corporate disclosure on credit risk valuation using news
coverage and news sentiment. Based on 13 years of CDS data, Tsai, Lu,
and Hung (2016) found a correlation between negative news sentiment and
increased credit risk.

Sentiment analysis uses dictionaries to classify the tone of the words
and then scores negative and positive word counts to measure the tone.
Loughran and Mcdonald (2009) shows that word lists developed for other
disciplines like Harvard Dictionary misclassify common words in the
financial text. In a large sample of 10-Ks from 1994 to 2008, almost
three-fourths of the terms identified as negative by the widely used
Harvard Dictionary are words typically not considered harmful in
financial contexts. Loughran and Mcdonald (2009) developed an
alternative negative word list and five other word lists that better
reflect tone in the financial text. They linked the word lists to 10 K
filing returns, trading volume, return volatility, fraud, material
weakness, and unexpected earnings. The research community has widely
accepted these dictionaries for financial text analysis. Some of the
other word lists used to measure the tone of financial are 1) a wordlist
developed in Henry's (2006, 2008) analysis of earnings announcements
(Henry Wordlist); 2) a wordlist from DICTION (DICTION Wordlist) software
developed by Roderick Hart; and 3) a wordlist from the General Inquirer
program (G.I. Wordlist) designed by social psychologist Philip Stone
(Henry and Leone (2014)).

Unlike other domains, positive and negative sentiments have asymmetric
information for investors. Due to their different implications for stock
prices, trading volumes, and firm fundamentals, netting positive and
negative sentiment measures results in significant information loss.
Azimi and Agrawal (2018) found that both negative and positive sentiment
measures explain variation in stock prices and trading volume at the
time of disclosure. These measures also have predictive power for future
profitability, cash holding, and leverage.

The textual sentiment of corporate disclosure is useful in forecasting
corporate investment and financing decisions. Firms whose tone in 10-k
filings is conservative are known to be risk-averse in their M\&A
actions. Ahmed and Elshandidy (2016) has concluded that firms with
conservative disclosure tone make conservative acquisition choices. They
are reluctant bidders, prefer stock-based acquisitions over cash-based
deals, and experience abnormal low stock returns during M\&A
announcements.

Financial disclosures risk sentiment has implications for future
earnings and stock returns. An increase in risk sentiment results in
lower future earnings, and risk sentiment explains future returns
variance. Liu et al. (2018) created and used 10-K risk-sentiment dataset
for risk detection. Li (2006) constructed risk sentiment measure of
annual reports using risk or uncertainty related word frequencies in the
10-K filings and tested the relation between the metric and firms'
future returns. They found that firms with a larger increase in risk
sentiment have more negative earnings changes in the next year. Also,
firms with increased risk sentiment underperform relative to other firms
twelve months after the annual report filing date. A risk
sentiment-based long-short portfolio generated a 10\% yearly Alpha,
measured using the four-factor model, i.e., momentum and the Fama-French
three factors. Gandhi, Loughran, and McDonald (2017) suggested using
annual report sentiment as a proxy for Financial Distress in U.S. banks.
They found that the annual report's negative sentiment is associated
with larger delisting probabilities. Sentiment also explains the odds of
paying subsequent dividends, loan loss provisions, and lower future
returns on assets.

Recent years have seen the usage of sentiment analysis as an input
feature to supervised learning techniques. Rawte, Gupta, and Zaki
(2018a) used sentiment analysis to predict bank failure. They combined
word sentiment polarities and their count as a weighted feature vector
for SVM and deep learning approaches. Tao, Deokar, and Deshmukh (2018b)
constructed numerous linguistics features from Forward-Looking
Statements (FLSs) like topics, sentiments, readability, semantic
similarity to analyze pre-IPO predictability vs.~post-IPO valuation.

\hypertarget{tone-analysis}{%
\subsubsection{Tone analysis}\label{tone-analysis}}

Tone analysis enquires whether the text content is optimistic or
pessimistic in tone. The tone of the forward-looking statements in a
firm's MDA explains future earnings and liquidity, even after
controlling other predictors of future performance. F. Li (2010b)
analyzed the relation between the tone of the forward-looking statements
(FLS) in the MDA and the firm's future returns. They manually
categorized 30,000 randomly selected FLS sentences on two dimensions
tone and content. Content is classified based on the topic of the
sentence, including profitability, liquidity, and operations. With this
as training data, they classified the tone and content of about 13
million forward-looking statements. They found that better current
performance and lower accruals are associated with higher positive
forward-looking statements in disclosures. A critical aspect of language
changes in usage also has information content. Feldman et al. (2008) and
Feldman et al. (2010) measured the change in disclosure tone compared to
the previous four filings and examined the incremental information
content of tone Change. Their results indicate that short window market
reactions around the SEC filing are associated with the MDA section's
tone, even after controlling for accruals, OCF, and earnings surprises.
The tone adds significantly to portfolio drift returns in the window of
time. They also found that incremental information of tone change is
more extensive for firms with weak information environment. Amel-Zadeh
and Faasse (2016) found that changes in the MDA text and footnotes and
differences between the two sections' tone predict negative future stock
returns and operating performance.

Others have examined the relationship between the tone of management's
10-K filings and the likelihood of getting involved in an FCPA violation
or litigation. Lopatta, Jaeschke, and Yi (2014) found that FCPA
violators use more negative, uncertain, litigious, and complex language
when disclosing financials than non-violators. They also show that
managers make strategic use of language after FCPA prosecution by
lowering their negative, uncertain, and litigious tone in 10-K filings.
On the other hand, investors who bought into stocks based on
management's optimistic projections may be disappointed by subsequent
underperformance. So, the use of optimistic language in disclosures may
increase litigation risk. Rogers, Van Buskirk, and Zechman (2011)
examined the relation between disclosure tone and shareholder
litigation. Based on tone measurements using general-purpose and
context-specific text dictionaries, they observed that sued firms'
earnings announcements are unusually optimistic relative to other firms
in similar economic circumstances. This observation indicates that
plaintiffs target more optimistic statements in their lawsuits.

In recent years, researchers have tried to study the relation between
tone and more complex business attributes. Researchers tried to explain
the business strategy, regulatory compliance, litigation risk, and
management characteristics in these works.

Manager's optimism or pessimism in disclosures can impact the market's
reaction to the information. Hendricks, Lang, and Merkley (2017) studied
the relation between the Manager's tone or optimism in disclosures and
subsequent firm performance. They observed founder-led firms have
``excess'' optimism relative to realized earnings and compared to
non-founder-led firms. Buchholz et al. (2018) examined the relation
between CEO narcissism and abnormal optimistic tone in financial
disclosures. They defined ``abnormal'' as the tone unrelated to a firm's
performance, risk, and complexity. In a sample of U.S. listed firms over
1992-2012, they observed that an abnormal, optimistic tone in 10-K
filings is positively related to CEO narcissism.

Mayew, Sethuraman, and Venkatachalam (2015) stressed the importance of
linguistic tone in assessing a firm's health. Using a sample of bankrupt
firms between 1995 and 2012, they concluded that management's opinion
about going-concern and the MDA's linguistic tone together predict
whether a firm will go bankrupt. F. Li (2010a) linked self-serving
attribution bias (``SAB'') to the Manager's tone in disclosures. FLS of
managers with more SAB have smaller variations in the tone (e.g.,
positive versus negative), and their earnings forecasts tend to be
overly optimistic. Firms whose managers have more SAB have higher
investment-cash flow sensitivity and experience more negative market
reactions around acquisition announcements.

Research results suggest that the tone ambiguity of a firm's financial
disclosures is related to managerial information hoarding. Shareholders
of firms with less readable and more ambiguous annual reports suffer
from a lack of transparency and bear the increased cost of external
financing. Ertugrul et al. (2016) investigated the impact of a firm's
annual report readability and ambiguous tone on its borrowing costs.
They find that firms with a higher proportion of uncertain and weak
modal words in 10-Ks have stricter loan contract terms and greater
future stock price crash risk. Lim, Chalmers, and Hanlon (2018) analyzed
the relationship between a firm's business strategy and the tone used in
disclosures. They find that prospectors display more negative and
uncertain styles, while defenders exhibit a more litigious manner in
their 10-Ks. Ji and Tan (2016) studied changes in firms' disclosure
policies in response to labor unemployment concern using the tone of 10K
and 10 Q filings and found it as an essential consideration for
corporate discretionary disclosure. Sandulescu (2015) investigated the
relations between disclosure tone, insider trading, and returns and
found that the net disclosure tone predicts the insider purchase ratio
(purchases scaled by the sum of purchases and sales) and abnormal
returns after controlling for past purchases, return volatility, and
firm characteristics.

Researchers also tested the validity of linguistic methods borrowed from
other domains. ``Diction'' is frequently used to assess the tone of
business documents. Loughran and McDonald (2015) argued that Diction is
not suitable for the same. More than two-thirds of the Diction
optimistic words and Diction pessimistic words in a sizeable 10-K sample
are likely misclassified. For example, words like respect, security,
power, and authority will not be considered positive by readers of
business documents. These are frequently in disclosures. Also, nearly
half the pessimistic 10-k word counts are ``not'' and ``no.'' The
authors compared Diction and Loughran-McDonald (2011) and believed that
the latter is better at capturing tone in the business text.

\hypertarget{readability}{%
\subsubsection{Readability}\label{readability}}

Readable disclosures and reports provide simple, straightforward, and
homogenous information that is understandable by all investors.
Readability measures the ease of understanding a text. Researchers
studied Financial disclosures readability for years and assessed its
impacts on various aspects of financial markets. Initial readability
research work focused on comparative analysis. Readability scores vary
between different sections of an annual report. Heath and Phelps (1984)
analyzed 20 randomly selected Fortune 500 companies, 1981 annual
reports, and nine business publications. They observed that financial
disclosure sections were difficult to read. At the same time, they found
that majority of business publications had similar readability scores.

Predominantly, annual reports and analyst's reports are the subjects of
the prior research work. The firm's annual report provides comprehensive
coverage of past achievements, enabling the readers to form, confirm, or
revise future performance expectations. For this, the report's content
must be readable and understandable. Report writers can improve
information communication by being responsive to their audiences'
reading and comprehension abilities (Courtis (1998)).

\emph{Readability measures}

One of the frequently used measure for readability is the FOG Index. It
is useful to differentiate school textbooks. It indicates the number of
years of education needed to understand the text on a first reading.
Thus, a Fog Index value of 16 implies that the reader needs 16 years of
schooling- essentially a college degree-to comprehend the text on a
first reading. It is a function of the average number of words in a
sentence (length) and multi-syllable words percentage. Words with more
than two syllables are challenging to comprehend.
\[Fog\: Index = 0.4 (average \:number\: of\: words\: per\: sentence +percentage\: of\: complex\: words)\]
One of the pioneering works that used the FOG index in disclosure
analysis on a large sample was by Li (2008). In this widely cited paper,
Li (2008) measured annual reports' readability using the Fog Index and
researched the relation with the firm's subsequent performance. Numerous
researchers have elaborated on this work and studied the association
between readability and other firm attributes.

However, the FOG index's applicability, whose focus is on grade texts,
has been questioned by other researchers. Using three different
readability measures on a sample of 42,357 10-Ks during 1994-2007,
Loughran and Mcdonald (2009) demonstrated that syllable counts for
assessing readability might not suit business applications. They argue
that of the Fog's two components, one is mis specified, and the other is
difficult to measure. Further, Loughran and McDonald (2014) reported
that file size of disclosure (10-K document) as a readability proxy
outperforms the Fog Index. This measure's advantages include eliminating
document parsing, replicability, and its correlation with other
readability measures.

\emph{Readability drivers}

While it is in analysts' and investors' interest to have readable
disclosures, management's incentives might not be aligned with the same.
Regulators took note of concerns about the complexity of firm
disclosures and initiated efforts to improve annual reports'
readability.

Regulators can influence the overall readability of disclosures with
rules and guidelines. In October 1998, the SEC-mandated that firms use
plain English in their prospectus and encouraged straightforward English
usage in disclosures. Based on textual analysis of Form 424, IPO
prospectus, and 10-K filings over 1994-2009, Loughran and McDonald
(2014), found that the SEC rule significantly impacted managers'
disclosure style. They also found that firms with better corporate
governance policies have higher compliance than others. Kubick and
Lockhart (2016) demonstrated that SEC oversight influences disclosure
practices and reduces the likelihood of stock price crashes. Similar
responses to regulation are observed in other developed economies. Smith
(2016) measured communication value using audit report readability and
the tone. They found that after the passage of ISA 700 (U.K. and
Ireland), audit reports are easier to read.

Li (2008) demonstrated that a firm's performance and readability of 10-K
filings have a statistically significant relation. Using the Fog Index
to measure readability, they find that loss-making firms use complex
sentences in disclosures, supporting the theory of management
obfuscation. Bloomfield (2008) tried to explain the drivers of length
and readability of annual reports by conducting a longitudinal study of
a single firm's 10-K's over three years.

Another critical factor that influences disclosure complexity and
readability is the firm's business complexity itself. As business
strategy fundamentally determines a firm's activities, it controls a
firm's operating complexity, environmental uncertainty, and information
asymmetry. Lim, Chalmers, and Hanlon (2018) investigated business
strategy as a determinant of annual report readability. They found that
firms pursuing an innovation-oriented prospector strategy have less
readable 10-Ks relative to firms seeking an efficiency-oriented defender
strategy. This finding suggests that a firm's strategy and operational
complexity must be considered while interpreting readability.

The above literature attributes readability metrics to the reporting
firms' operational complexity and obfuscation attempts. Contrary to
this, Xu, Fernando, and Tam (2018) tried to explain the readability with
the help of management age. Using upper echelons theory and business and
social science research, they suggested that older CEOs and executives
may be better at explaining operational complexities and staying ethical
in reporting, thus leading to more readable financial reports.

\hypertarget{readability-vs-firm-features}{%
\paragraph{Readability vs firm
features}\label{readability-vs-firm-features}}

As the prior sub-section showed, readability has been used extensively
in financial text analysis. The next subsection evaluates various firm
attributes explained by readability measures in the literature. Firm
profitability is one of the key measures that drive a firm's valuation.
Analysts and investors are interested in explaining and forecasting
variations in profitability measures and subsequent market returns. The
initial focus of text analysis and readability of disclosures were on
this topic. Li (2008) demonstrated that the readability of 10-K filings
has a statistically significant impact on a firm's subsequent
performance. They find that firms with losses, or with transient income,
write annual reports with long sentences and big words. On the contrary,
Lo, Ramos, and Rogo (2017) found a relationship between a firm's ability
to beat the prior year's earnings and disclosure complexity. This
reduction in readability with an increase in profits is contrary to Li
(2008) findings and questions the assumption that good news is easier to
communicate.

The participation of a large number of investors and specifically, small
investors is essential for efficient markets. One measure that reflects
this broad participation is trading liquidity. Investors' may find it
challenging to understand and analyze complex disclosures and annual
reports, which can reduce their willingness to trade, decreasing stock
liquidity. Loughran and Mcdonald (2009) found a significant relation
between improved 10-K readability and increased small investor trading
as well as the likelihood of seasoned equity issuance. Studying market
response to SEC EDGAR 10-K filings You and Zhang (2009), found that
complex 10-K filings result in investors' under-reaction. Boubaker,
Gounopoulos, and Rjiba (2019) examined the effect of annual report
textual complexity on firms' stock liquidity and found that lower
readability is related to more inferior stock liquidity. Their findings
were robust to sensitivity tests, including endogeneity, alternative
estimation techniques, and alternative liquidity and readability
proxies.

If investors understand the annual reports, the stock price should
respond to their information content. If investors cannot understand
firm-specific information, the firm stock price may move broadly in line
with the market. Stock return synchronicity, or co-movement, measures
the extent to which the individual stock returns would comove with
market returns. Improved readability in financial reports may reduce
firm-specific information-processing costs and, therefore, minimize
stock return synchronicity. Bai, Dong, and Hu (2019) investigated the
relation between firm-specific information-processing cost, proxied by
annual report readability, and investors' firm-specific information
usage, proxied by firm stock return synchronicity. They found that
annual reports' readability is negatively related to the firm's future
stock return synchronicity. They note that this relation is more
concentrated on firms with low analyst coverage or institutional
ownership. As the impact is more focused in a specific type of firms-
small firms, firms with high R\&D spending, or high growth firms, they
relate it to information asymmetry

Another side of information asymmetry is management information
hoarding. Managements may not be willing to share the information
adequately for various reasons. Firms with less readable annual reports
suffer from less transparent information disclosure and may bear the
increased cost of external capital as lenders might charge a premium due
to uncertainty. In their investigation of the relation between a firm's
annual report readability and borrowing costs, Ertugrul et al. (2016),
firms with lower readability scores and higher uncertain tone in 10-Ks
have stricter loan contract terms and greater future stock price crash
risk. Another significant aspect of a firm's behavior that investors and
analysts are worried about is the possibility of on-going fraud.
Obfuscation and deception theories from accounting and communication
literature suggest that management producing 10-Ks can deceive by hiding
harmful news in complex and unreadable content while highlighting
achievements. Moffitt and Burns (2009) examined 202 fraudulent and
non-fraudulent 10-Ks by focusing on 25 linguistic cues. The results
indicate that fraudulent 10-Ks have more complex words, signaling words
of achievement and cause, and qualifying conjunctions. They also noted
that truthful 10-Ks have better FRE readability and use more present
tense verbs. Others have studied firms indulging in Foreign Corrupt
Practices. Lopatta, Jaeschke, and Yi (2014) tried to detect a firm's
likelihood of violating the FCPA using disclosure tone and readability.
Management of FCPA violators tends to use negative, uncertain,
litigious, and complex language than non-violators. They also noted that
managers strategically adjust tone in 10-K filings after FCPA
prosecution. One consequence of fraud is that investors can take legal
recourse if they incur a financial loss due to the misleading
information shared by management in financial disclosures. So, the text
content and narrative disclosures can influence subsequent shareholder
litigation. Based on federal securities class action lawsuits over two
decades, propensity-score matched samples, and linguistic measures
(readability and sentiments) in textual disclosures, Ganguly (2017),
found poor readability in disclosure is often perceived as misleading
after the fact and results in litigations.

Other researchers have explored the importance of numbers in
understanding financial disclosures and questioned their exclusion in
prevalent textual analysis research. Siano and Wysocki (2018) showed
that the prevalence of numbers in annual reports is positively
correlated with the disclosure's readability. They also demonstrate that
prior findings on the links between disclosure readability and firm
performance can be explained by the ubiquity of numbers in the
disclosures. These studies limit, if not undermine, the applicability of
readability research.

\emph{Analyst reaction}

Prior research focused on investors and analysts as investment
decision-makers and how they use disclosures. Managers use some of the
disclosure content to influence analysts as facilitators. As analysts
act as intermediaries between management and final investors, the
analyst needs to comprehend management's signal. Also, as analysts track
specific industries, industrial jargon in disclosure would not deter
them and will not hinder their ability to understand disclosures. In
recent years, academic research in this domain has picked up. Based on
10-k readability, Lehavy, Li, and Merkley (2011) documented that less
readable reports require analysts to spend more effort to generate
reports. They find that less readable 10-Ks are associated with higher
uncertainty in analyst earnings forecasts. Their findings are in line
with the expectation that more analyst services are required for
evaluating firms with lesser readability content in disclosures

Readability reduces the agency costs and information asymmetry between
investors, attracting more financial analysts to track a firm. Diaz,
Njoroge, and Shane (2017) suggests that increased uncertainty created by
greater complexity, opacity, and volume of information in 10-K filings
causes analysts to react with more restraint compared to the time of
earnings announcements. Sourour, Badreddine, and Aymen (2018), using the
Gunning Fog index and the Flesh Reading Ease formula, investigated 88
companies listed on the French CAC All between 2009 and 2014. They
concluded that number of analysts following a firm, and as a result, the
attention paid by institutional investors is directly related to the
readability of the financial disclosures. The recommended using short
sentences, familiar words, or the active voice in disclosures reduces
the cognitive distance between management and investors.

Other researchers have studied the relation between analysts' reports
readability and stock price performance as it may affect value-relevant
information. Hsieh and Hui (2013) found that analyst report readability
reduces forecast dispersion, and that market reaction is significantly
positive towards more readable reports. Hsieh, Hui, and Zhang (2016)
further finds that this effect is significantly positive only for firms
with information asymmetry, i.e., higher R\&D spending, higher bid-ask
spreads, and a greater proportion of uninformed investors, and more
experienced analysts. These empirical findings indicate that analysts'
reports reduce future earnings uncertainty, and investors incorporate it
into the market price.

This sub-chapter has surveyed how researchers have conducted financial
text analysis. The next section will review various organizational
outcomes that researchers have tried to explain using disclosure text
content.

\hypertarget{firm-specific-attributes-and-events-analysis}{%
\section{Firm-Specific attributes and events
analysis}\label{firm-specific-attributes-and-events-analysis}}

This sub-chapter explores firm attributes that researchers have studied
using disclosure content. A public firm is an ongoing concern and goes
through different stages in its life cycle. Firms respond to and
influence the drivers in the ecosystem through planned and reactive
measures. This behavior results in measurable changes in firm
attributes. Investors are interested in target company attributes like
business strategy, exposure to risk, senior management performance,
fraudulent reporting, competition, profitability, and financial
distress. Researchers have found evidence linking many firm attributes
and the language used in disclosures. One of the foremost attributes of
a firm is its business strategy. As business strategy fundamentally
determines a firm's product and market domain, technology, and
organizational structure, it influences a firm's operating complexity,
environmental uncertainty, and information asymmetry. Consequently, the
business strategy frames the level, wording, and complexity of
disclosures. Lim, Chalmers, and Hanlon (2018) investigated the influence
of business strategy on annual report readability. They measured 10-K
readability with Li's (2008) Fog index. They found that firms pursuing
an innovation-oriented prospector strategy have less readable 10-Ks
relative to firms seeking an efficiency-oriented defender strategy. They
also found that prospectors display more negative and uncertain tones
while defenders exhibit a more litigious tone in their 10-Ks. Fengli,
Lundholm, and Michael (2013) observed limited and indirect evidence that
management strategically makes misleading statements about their
competitive landscape. Bushman, Hendricks, and Williams (2014) used
textual analysis to extract a bank-specific competition measure to
examine the relationship between competition and bank stability. CEO
characteristics influence a firm's performance and disclosure
narratives. Hendricks, Lang, and Merkley (2017) examined whether firms'
regulatory filings' textual attributes reflect CEO characteristics and
whether investors consider this relation when assessing firm value. They
found that 10-K text for founder-led firms has ``excess'' optimism
relative to current and future realized earnings and compared to
non-founder-led firms. They provided broad sample evidence that CEO
fixed effects are significantly related to several textual attributes.
Subsequently, Buchholz et al. (2018) examined the driver behind this
excessive over-optimism and linked it to CEO narcissism

\hypertarget{ethics-trust}{%
\subsubsection{Ethics \& trust}\label{ethics-trust}}

With the increased importance of corporates role in society,
stakeholders expect firms to conduct their business in an ethical and
trustworthy manner. It is challenging to measure Ethical behavior using
traditional financial metrics. Trust is ethically important and
essential for business. It is not easy to measure whether a firm has a
trusting corporate culture. Audi, Loughran, and McDonald (2016)
developed an objective measure of trust in a firm's corporate culture by
counting the frequency of 21 trust-related words in MDA. They concluded
that firms with a trusting culture frequently use audit and control-type
words and that trust explains subsequent share price volatility.
Loughran and Mcdonald (2016) examined the occurrence of ethics-related
terms in 10-K annual reports. They observed that ethics-related terms
are frequently used by ``sin'' stocks. These firms have poor corporate
governance and are likely to be the target of lawsuits.

\hypertarget{financial-distress}{%
\subsection{Financial distress}\label{financial-distress}}

Investors are keen on knowing about the health of the firms they may
invest in the future. A firm in financial distress loses a significant
amount of its shareholder's value. If the management cannot tide over
the crisis, the firm may have to file for bankruptcy, resulting in a
50\% to 80\% loss of capital for shareholders and lenders. Financial
distress and bankruptcy prediction is an actively researched field.
Researchers attempted to incorporate text features into such predictive
models. Below is a brief review of the same. Auditors express
going-concern opinions based on the firm's obligations and liquidity.
Financial disclosures include these opinions. Change in such disclosures
can act as a signal to identify distress. However, auditors do respond
to external financial markets. Beams and Yan (2015) examined the
financial crisis's effect on auditor going-concern opinions and
concluded that the financial crisis led to increased auditor
conservatism. A going-concern opinion in disclosures is associated with
the number of forward-looking disclosures and their ambiguity. Enev
(2017) observed that while the absolute number of forward-looking
disclosures is lower for companies receiving a going concern opinion,
the proportion of forward-looking disclosures in the MDA is higher in
the presence of a going concern opinion. The results suggest generally
improved forward-looking disclosures in MDA when companies receive a
going concern opinion from their auditor. One consequence of distress is
financial constraints. Firms undergo reduced cash flows during stress,
which results in liquidity events - like dividend omissions or
increases, equity recycling, and underfunded pensions. Analysts measure
the extent of financial constraints to assess the capital structure.
Bodnaruk, Loughran, and McDonald (2013) used constraining-words based
lexicon to measure the same. These measures have a low correlation with
traditional financial constraints measures and predict subsequent
liquidity events better. Ball, Hoberg, and Maksimovic (2012) used text
in firms' 10-Ks to measure investment delays due to financial
constraints. They found that the fundamental limitations are the
financing of R\&D expenditures rather than capital expenditures and that
the main challenge for firms is raising equity capital to fund growth
opportunities. These text-based measures predict investment cuts
following the financial crisis better than other indices of financial
constraints used in the literature.

\hypertarget{bankruptcy}{%
\subsection{Bankruptcy}\label{bankruptcy}}

Once a company is unable to come out of distress, it will become
insolvent. Insolvency is the state in which the company is not capable
of honoring some commitment. Lenders and claim holders can force the
insolvent company to discontinue operations. Managements file for
bankruptcy protection to recover from such a situation or liquidate it
in an orderly manner. Bankruptcy prediction has been an active research
topic for accounting researchers over decades. Most prior bankruptcy
prediction models were developed by using financial ratios. However,
signs of distress may appear in the nonfinancial information earlier
than changes in the financial ratios. Current distress measures tend to
miss extreme events, especially in the banking sector (Gandhi, Loughran,
and McDonald (2017)). In recent years, qualitative information and text
analysis have become necessary because frequent changes in accounting
standards have made it difficult to compare financial numbers between
years (Shirata et al. (2011)). The language used by future bankrupt
companies differs from non-bankrupt companies. Hájek and Olej (2015)
studied various word categories from corporate annual reports and showed
that the language used by bankrupt companies shows stronger tenacity,
accomplishment, familiarity, present concern, exclusion, and denial.
Bankrupt companies also use more modal, positive, uncertain, and
negative language. They built prediction models combining both financial
indicators and word categorizations as input variables. This
differential language usage is observed in non-English firms'
disclosures also. Shirata et al. (2011) analyzed the sentences in
Japanese financial reports to predict bankruptcy. Their research
revealed that co-occurrence of words ``dividend'' or ``retained
earnings'' in a section distinguish between bankrupt companies and
non-bankrupt companies.

Working on U.S. Banks Gandhi, Loughran, and McDonald (2017) used
disclosure text sentiment as a proxy for bank distress. They found that
the annual report's more negative sentiment is associated with larger
delisting probabilities, lower odds of paying subsequent dividends,
higher subsequent loan loss provisions, and lower future return on
assets. Similarly, Lopatta, Gloger, and Jaeschke (2017) concluded that
firms at risk of bankruptcy use significantly more negative words in
their 10-K filings than comparable vital companies. This relationship
holds up until three years before the actual bankruptcy filing. Other
notable works using text analysis for bankruptcy prediction were Yang,
Dolar, and Mo (2018) and Mayew, Sethuraman, and Venkatachalam (2015).
Yang, Dolar, and Mo (2018) used high-frequency words from MDA and
compared the differences between bankrupt and non-bankrupt companies.
Mayew, Sethuraman, and Venkatachalam (2015) also analyzed MDA with a
focus on going-concern options. They found that both management's
opinion about going concern reported in the MDA and the MDA's linguistic
tone together provide significant explanatory power in predicting
whether a firm will cease as a going concern. Also, the predictive
ability of disclosure is incremental to financial ratios, market-based
variables, even the auditor's going concern opinion and extends to three
years before the bankruptcy.

This sub-chapter reviewed the firm attributes that researchers tried to
explain using disclosure narratives. The majority of the work focused on
explaining risk exposure, fraud, financial distress, and bankruptcy. Of
these, default is a significant event in a firm's life and significantly
impacts stakeholders. Hence this research focuses on the bankruptcy
prediction task. The next sub-chapter will review the models used by
analysts for text analysis.

\hypertarget{modeling-approaches-used}{%
\section{Modeling approaches used}\label{modeling-approaches-used}}

The previous sections reviewed text analysis in finance in terms of the
content analyzed, linguistic features extracted from the text, and firm
attributes explained with those features. This section outlines the
statistical methodologies and models used in the text analysis of
financial disclosures. There are two dimensions to the models in Text
analysis in finance. First deals with the language model---the other
deals with underlying economic or financial process modeling.

\hypertarget{language-models}{%
\subsection{Language models}\label{language-models}}

As natural language is messy, researchers convert information from texts
into quantifiable variables and then use them in subsequent modeling
efforts. Language models make assumptions about how information is
encoded in language and translate it into a usable form. There are
broadly two categories of language models Dictionary-based and
statistical models. While they try to capture information in a language
differently, these models are complementary (Matthies and Coners (2015))

In dictionary-based models, the text under consideration is split into
words, and the words are categorized based on dictionaries. One can
consider a document in terms of raw word frequencies in it. This method
would result in a large number of dimensions, as every word in the
dictionary has representation. Context or domain-specific dictionaries
are useful in reducing the dimensionality compared to raw word
frequencies. Some of the dictionaries used are Diction, General
Inquirer, and the Linguistic Inquiry and Word Count (F. Li (2010b)),
Harvard and Loughran. Loughran built their dictionary explicitly for
text analysis in finance. Li (2006) measured risk-sentiment of 10-k
disclosures using risk or uncertainty related word frequency.

Dictionary-based models suffer from the shortcomings of all Bag of Words
approaches. These models consider words as the critical information
block and ignore the order of words. Also, any misclassification in the
dictionary will lead to erroneous results Loughran and Mcdonald (2009).
Loughran and Mcdonald (2009), Loughran and McDonald (2014), discuss how
usage of generic dictionaries leads to large scale miss-classification
of words. Also, Van Den Bogaerd and Aerts (2011) noted that while most
of the text classification work in the accounting industry and research
is done manually, an erroneous, expensive process, few research papers
mention the accuracy of the used classification methodology.

Another challenge with raw word frequency is frequently occurring words,
which might not have essential information. Alternatively, one can use
novelty-based weights. Term frequency- Inverse Document Frequency
(TF-IDF) gives higher weights to frequent terms that are infrequent
across the documents. Smailović et al. (2018) conducted a differential
content analysis based on TF-IDF weighting and evaluated correlation
with financial variables. They considered linguistic features such as
personal/impersonal pronouns ratio, sentiment, trust, doubt, certainty,
and modality. Qiu (2007) constructed TF-IDF from financial statements
and built SVM-based classifiers to predict future company performance.

In recent years, continuous or vector representation of words has
dramatically improved NLP tasks performance in other domains. Some of
the earlier approaches include latent representations like Latent
Semantic Analysis (LSA) and Latent Dirichlet Analysis (LDA). Later
methods include Word2Vec, Global Vectors (GloVE), Contextual word
vectors (CoVE), and Elmo. These were predominantly encoder driven
architectures. In the past two years transformer driven language
representations have resulted in state-of-the-art performance on
numerous NLP tasks. Some of them are Bidirectional Encoder
Representations from Transformers (BERT), Generative Pre-Training (GPT1,
and GPT2). While researchers used probabilistic distribution language
models like LDA in prior research (Chen et al. (2018)), they are yet to
use other advanced language models.

\hypertarget{process-models}{%
\subsection{Process models}\label{process-models}}

Research in Financial text analysis is motivated by the need to
understand firms' disclosure information and investors' sentiments about
firms. The availability of public disclosures shapes much of the
research work in the domain. As most of the studies were correlational,
analysts constructed various regression models based on their objective.
While previous years have seen the dominance of hand-crafted variables
from the text in financial disclosures, representative learning, and
deep learning models have emerged as an alternative in the past 24
months.

\hypertarget{regression-models}{%
\subsubsection{Regression models}\label{regression-models}}

Regression models are useful in analyzing variables in finance and
accounting. Researchers extended the same approach to incorporate text
information. The text features could be raw word frequencies, derived
features based on dictionaries, or outputs of prediction and ranking
models that transform some firm attributes into numeric form. Foster and
Hussey (2006) studied corporate transformation and its success factors
in large U.S. listed firms. Selecting firms with a deteriorating
performance that may need strategic shifts, they built a prediction
model to identify formalized transformation programs based on
restructuring costs and significant corporate announcements. They
extracted the firm's strategic intent from annual reports MDA and
converted it to a numerical value. Using these features, they
constructed a regression model to explain the relationship between the
number of factors on change in total shareholder return (TSR) during
transformations. Bochkay (2014) used the ``bag-of-words'' (BOW) approach
to represent MD\&A sections and regularized regression methods to handle
high-dimensionality and multicollinearity. Others conducted correlation
studies predominantly to understand the relations between linguistic
features and financial aspects Smailović et al. (2018).

\hypertarget{bayesian-models}{%
\subsubsection{Bayesian models}\label{bayesian-models}}

One of the preferred statistical modeling approaches is Naive Bayesian
machine learning. F. Li (2010b) assessed the information content of the
forward-looking statements (FLS) in MDA of 10-K and 10-Q filings using a
Naive Bayesian machine learning algorithm. Brown, Crowley, and Elliott
(2016) used a Bayesian topic modeling algorithm to quantify the topic
content of annual reports and assessed if it is incrementally
informative in predicting intentional misreporting. Humpherys et
al.~used Naive Bayes and decision trees on public disclosures for
deception detection. For text sequence modeling, conditional Random
Fields (CRF), a discrete classifier model, has been used. Chen et al.
(2011), chen (2013), extracted opinioned statements from MDA of 10-K,
using CRF based tagging models. They considered multiple combinations of
linguistic factors such as predicate-argument structure, morphology,
syntax, orthography, and simple semantics.

\hypertarget{comparative-analysis}{%
\subsubsection{Comparative analysis}\label{comparative-analysis}}

While individual approaches have their merits, comparative analysis in
the given context can throw light on models' strengths. Chen et al.
(2018) used topic modeling and evaluated multiple topic modeling
approaches and their effectiveness. They applied four primary topic
modeling methods - Principal Component Analysis, Non-negative Matrix
Factorization, Latent Dirichlet Allocation, and KATE, to analyze SEC
filings by U.S. public banks. Others have used ranking, clustering, or
SVM models to explain the processes. Tsai and Wang (2013), Qian and Li
(2013), attempted to use text in disclosures to rank the risk levels of
a set of companies. Glancy and Yadav analyzed MDAs in 10-K filings for
deceptive statements, using expectation maximization and hierarchical
clustering. Qiu (2007) built SVM based classifiers to predict future
company performance based on linguistic features.

\hypertarget{deep-learning}{%
\subsection{Deep learning}\label{deep-learning}}

Over the years, in classification tasks, logistic regression models have
outperformed other fraud detection models and other problems of
quantitative financial analysis (Beams and Yan (2015), Bose, Piramuthu,
and Shaw (2011), LeWitt (1988), Zhang (2016), Mousa Albashrawi (2016),
Bao et al. (2015)). In his evaluation of Machine Learning models for
Fraud detection, Zhang (2016) compares five state-of-the-art
classifiers, i.e., logistic regression, artificial neural network,
support vector machines, decision trees, and bagging, and concludes that
bagging performs the best. The non-linear nature of the underlying
process can explain this outperformance. Hence, neural network models,
which are viewed as a series of logits, can excel in these tasks. The
advantages of the neural network models have motivated some researchers
to use them in text analysis in finance. In a review of various text
analysis methods, Guo, Shi, and Tu (2017), note that neural network
outperforms other machine learning methods in news category
classification.

Qualitative information in corporate annual reports is more abundant
than what the previous models have been able to capture and leverage.
When measured actively, many of the linguistic features are informative
(Azimi and Agrawal (2018)). With improved hardware performance and
better algorithms, multi-layer neural networks have become economical to
train in recent years. This method, also known as deep learning, has
resulted in state-of-the-art performance in many NLP tasks. Encouraged
by the trend, some researchers have tried to use deep learning in text
analysis of finance. Tao, Deokar, and Deshmukh (2018a) used deep
learning methods to predict pre-IPO price revisions and post-IPO
first-day returns using forward-looking statements (FLSs).

While traditional regression models yield good results while using
limited hardware, the out of sample results are not satisfactory. The
advent of deep learning methods led to researchers exploring them in the
text analysis of disclosures. Azimi and Agrawal (2018) used deep
learning to measure and classify sentiment in SEC 10-K filings and
achieved 90\% accuracy in out-of-sample tests. These sentiment
indicators have a better predictive power of future stock performance.
Similarly, Song (2017) extracted textual features from Item 1 Business
descriptions in 10-K, using a deep learning algorithm to incorporate
word contexts. Rawte, Gupta, and Zaki (2018b) used deep learning and
Support Vector Machines (SVM) for bank failure classification using
textual analysis. Using changes in the 1A sections, they built models to
predict the risk indicators like leverage and Return On Assets (ROA).
The above findings indicate that while deep learning methods help
extract more information or signal from narrative disclosures, they are
yet to be used extensively in text analysis of finance.

\hypertarget{conclusion}{%
\section{Conclusion}\label{conclusion}}

Prior sections lay out the need for better tools to extract knowledge
from the financial disclosure test. Research in financial text analysis
used disclosure text metrics such as readability and sentiment to
explain attributes like risk exposure and financial distress. These
works used text features along with traditional accounting metrics.
These metrics, while having meaningful signals, discard most of the
information in the disclosure text. Hence the marginal gain in the
knowledge from such partial text analysis may not justify the efforts
required for undertaking financial text analysis.

Narrative content in financial disclosures has critical information
required to understand the organization state, predicting future
organizational outcomes. Extracting this information and identifying the
relationship between such information and organizational outcomes is
also critical for investors' ability to use the same.

This work has surveyed the literature on text analysis in finance,
explored frequently applied methods, and delineated the components that
have been the focus of studies. The following salient points emerge from
this survey of the prior work:

\begin{enumerate}
\def\labelenumi{\arabic{enumi})}
\item
  Text analysis in finance is still in the early stages, focusing on
  broad information extraction indicators, like readability, FOG index,
  and sentiment analysis.
\item
  Readability-driven text analysis may not be suitable for measuring the
  effectiveness of business communication. Loughran and McDonald (2014)
  showed that financial jargon and information integration are
  correlated.
\item
  Generic dictionary-based methods suffer from misclassification issues.
\item
  Many research questions can be answered with proper text analysis of
  financial disclosure documents
\end{enumerate}

\begin{itemize}
\tightlist
\item
  Firms' financial health, stress, and future bankruptcy can be assessed
  from ongoing concern comments.
\item
  Fraud and misreporting can be assessed from management disclosures and
  notes to accounts.
\item
  Companies future litigation risk, operational risk, and other risks
  can be assessed from risk disclosures
\item
  Firm's business strategy, competitive position, and strategy changes
  can be captured from product and business descriptions.
\end{itemize}

\begin{enumerate}
\def\labelenumi{\arabic{enumi})}
\setcounter{enumi}{4}
\item
  Efficient extraction of information from the text can enhance
  narrative texts' forecasting power about firm attributes.
\item
  Machine learning methods and neural representation learning methods
  can be useful in financial text analytics.
\end{enumerate}

\hypertarget{future-research}{%
\subsection{Future research}\label{future-research}}

Investors need computational methods to extract knowledge from
disclosure texts that can explain firm attributes. Researchers need to
examine the information contained in the financial disclosures text and
its relationship to organizational outcomes.

The research question can be framed as below.

\begin{itemize}
\tightlist
\item
  Given a firm's financial disclosure, what can we infer about its
  current state?\\
\item
  Which of the possible organizational outcomes can we predict?
\end{itemize}

From a methodology perspective, the research question can be framed as
follows. Given a prior labeled dataset of firm outcomes and relevant
period disclosure documents,

\begin{itemize}
\tightlist
\item
  Which of the textual features of the disclosure are relevant for
  analysis?
\item
  Can we infer relations between text features and organizational
  outcomes?
\item
  Can we build predictive models?
\end{itemize}

The following research work is required to answer the above questions.

\begin{itemize}
\item
  Conduct quantitative correlation studies to identify relations between
  different linguistic attributes and organizational outcomes.\\
\item
  Considering the information in text disclosures, exclusively
  text-based analysis methods can be feasible and valuable in financial
  and accounting analysis. Build predictive and classification models
  purely based on linguistic disclosure features.
\item
  Domain-specific dictionary methods can overcome generic dictionary
  methods in text analysis. Build multi-task and task-specific
  dictionary models and evaluate their predictive and explanatory power.
\item
  Modern representation learning techniques and affordable hardware and
  open source tools allow large scale machine learning. These tools help
  in overcoming the limitations of ``bag of words'' approaches. Build
  representative language models for financial disclosure text analysis
  and evaluate their performance.
\end{itemize}

\clearpage

\hypertarget{refs}{}
\begin{cslreferences}
\leavevmode\hypertarget{ref-Ahmed2016}{}%
Ahmed, Yousry, and Tamer Elshandidy. 2016. ``The effect of bidder
conservatism on M\&A decisions: Text-based evidence from US 10-K
filings.'' \emph{International Review of Financial Analysis} 46:
176--90. \url{https://doi.org/10.1016/j.irfa.2016.05.006}.

\leavevmode\hypertarget{ref-altman1968financial}{}%
Altman, Edward I. 1968. ``Financial Ratios, Discriminant Analysis and
the Prediction of Corporate Bankruptcy.'' \emph{The Journal of Finance}
23 (4): 589--609.

\leavevmode\hypertarget{ref-Amel-Zadeh2016}{}%
Amel-Zadeh, Amir, and Jonathan Faasse. 2016. ``The Information Content
of 10-K Narratives: Comparing MD\&A and Footnotes Disclosures.''
\url{https://doi.org/10.2139/ssrn.2807546}.

\leavevmode\hypertarget{ref-Asthana2001}{}%
Asthana, Sharad, and Steven Balsam. 2001. ``The effect of EDGAR on the
market reaction to 10-K filings.'' \emph{Journal of Accounting and
Public Policy} 20 (4-5): 349--72.
\url{https://doi.org/10.1016/S0278-4254(01)00035-7}.

\leavevmode\hypertarget{ref-Audi2016}{}%
Audi, Robert, Tim Loughran, and Bill McDonald. 2016. ``Trust, but
Verify: MD\&A Language and the Role of Trust in Corporate Culture.''
\emph{Journal of Business Ethics} 139 (3): 551--61.
\url{https://doi.org/10.1007/s10551-015-2659-4}.

\leavevmode\hypertarget{ref-Azimi2018}{}%
Azimi, Mehran, and Anup Agrawal. 2018. ``Is the Sentiment in Corporate
Annual Reports Informative? Evidence from Deep Learning.''
\url{https://doi.org/10.2139/ssrn.3258821}.

\leavevmode\hypertarget{ref-Bai2019}{}%
Bai, Xuelian, Yi Dong, and Nan Hu. 2019. ``Financial report readability
and stock return synchronicity.'' \emph{Applied Economics} 51 (4):
346--63. \url{https://doi.org/10.1080/00036846.2018.1495824}.

\leavevmode\hypertarget{ref-Ball2012}{}%
Ball, Christopher, Gerard Hoberg, and Vojislav Maksimovic. 2012.
``Redefining Financial Constraints: A Text-Based Analysis.'' \emph{SSRN
Electronic Journal}. \url{https://doi.org/10.2139/ssrn.1923467}.

\leavevmode\hypertarget{ref-Bao2015}{}%
Bao, Yang, Bin Ke, Bin Li, Y. Julia Yu, and Jie Zhang. 2015. ``Detecting
Accounting Frauds in Publicly Traded U.S. Firms: New Perspective and New
Method.'' \emph{Ssrn}, 173--88.
\url{https://doi.org/10.2139/ssrn.2670703}.

\leavevmode\hypertarget{ref-Bartley2011}{}%
Bartley, Jon, Al Y S Chen, and Eileen Z. Taylor. 2011. ``A comparison of
XBRL filings to corporate 10-Ks-evidence from the voluntary filing
program.'' \emph{Accounting Horizons} 25 (2): 227--45.
\url{https://doi.org/10.2308/acch-10028}.

\leavevmode\hypertarget{ref-Beams2015}{}%
Beams, Joseph, and Yun Chia Yan. 2015. ``The effect of financial crisis
on auditor conservatism: US evidence.'' \emph{Accounting Research
Journal} 28 (2): 160--71.
\url{https://doi.org/10.1108/ARJ-06-2013-0033}.

\leavevmode\hypertarget{ref-Ben-Rephael2017}{}%
Ben-Rephael, Azi, Zhi Da, Peter D. Easton, and Ryan D. Israelsen. 2017.
``Who Pays Attention to SEC Form 8-K?'' \emph{Ssrn}.
\url{https://doi.org/10.2139/ssrn.2942503}.

\leavevmode\hypertarget{ref-Blankespoor2014}{}%
Blankespoor, Elizabeth, Brian P. Miller, and Hal D. White. 2014.
``Initial evidence on the market impact of the XBRL mandate.''
\emph{Review of Accounting Studies} 19 (4): 1468--1503.
\url{https://doi.org/10.1007/s11142-013-9273-4}.

\leavevmode\hypertarget{ref-Bloomfield2008}{}%
Bloomfield, Robert. 2008. ``Discussion of "Annual report readability,
current earnings, and earnings persistence".'' \emph{Journal of
Accounting and Economics} 45 (2-3): 248--52.
\url{https://doi.org/10.1016/j.jacceco.2008.04.002}.

\leavevmode\hypertarget{ref-Bochkay2014a}{}%
Bochkay, B Y Khrystyna. 2014. ``Enhancing Empirical Accounting Models
with Textual Information.''
\url{https://rucore.libraries.rutgers.edu/rutgers-lib/43748/PDF/1/play/}.

\leavevmode\hypertarget{ref-Bodnaruk2013}{}%
Bodnaruk, Andriy, Tim Loughran, and Bill McDonald. 2013. ``Using 10-K
Text to Gauge Financial Constraints.'' \emph{Ssrn} 50 (4): 623--46.
\url{https://doi.org/10.2139/ssrn.2331544}.

\leavevmode\hypertarget{ref-Bose2011}{}%
Bose, Indranil, Selwyn Piramuthu, and Michael J. Shaw. 2011.
``Quantitative methods for Detection of Financial Fraud.''
\emph{Decision Support Systems} 50 (3): 557--58.
\url{https://doi.org/10.1016/j.dss.2010.08.005}.

\leavevmode\hypertarget{ref-Boubaker2019}{}%
Boubaker, Sabri, Dimitrios Gounopoulos, and Hatem Rjiba. 2019. ``Annual
report readability and stock liquidity.'' \emph{Financial Markets,
Institutions and Instruments}, 41.
\url{https://doi.org/10.1111/fmii.12110}.

\leavevmode\hypertarget{ref-Bourveau2018}{}%
Bourveau, Thomas, Yun Lou, and Rencheng Wang. 2018. ``Shareholder
Litigation and Corporate Disclosure: Evidence from Derivative
Lawsuits.'' \emph{Journal of Accounting Research} 56 (3): 797--842.
\url{https://doi.org/10.1111/1475-679X.12191}.

\leavevmode\hypertarget{ref-Brown2016}{}%
Brown, Nerissa C., Richard Crowley, and W. Brooke Elliott. 2016. ``What
are You Saying? Using Topic to Detect Financial Misreporting.''
\emph{Ssrn}, March. \url{https://doi.org/10.2139/ssrn.2803733}.

\leavevmode\hypertarget{ref-Brown2011}{}%
Brown, Stephen V., and Jennifer Wu Tucker. 2011. ``Large-Sample Evidence
on Firms' Year-over-Year MD\&A Modifications.'' \emph{Journal of
Accounting Research} 49 (2): 309--46.
\url{https://doi.org/10.1111/j.1475-679X.2010.00396.x}.

\leavevmode\hypertarget{ref-Buchholz2018}{}%
Buchholz, Frerich, Reemda Jaeschke, Kerstin Lopatta, and Karen Maas.
2018. ``The use of optimistic tone by narcissistic CEOs.''
\emph{Accounting, Auditing and Accountability Journal} 31 (2): 531--62.
\url{https://doi.org/10.1108/AAAJ-11-2015-2292}.

\leavevmode\hypertarget{ref-Bushman2014}{}%
Bushman, Robert M., Bradley E. Hendricks, and Christopher D. Williams.
2014. ``The Effect of Bank Competition on Accounting Choices,
Operational Decisions and Bank Stability: A Text Based Analysis.''
\emph{SSRN Electronic Journal}.
\url{https://doi.org/10.2139/ssrn.2460371}.

\leavevmode\hypertarget{ref-Buhner1985}{}%
Bühner, R., and P. Möller. 1985. ``the Information Content of Corporate
Disclosures of Divisionalization Decisions {[}1{]}.'' \emph{Journal of
Management Studies} 22 (3): 309--26.
\url{https://doi.org/10.1111/j.1467-6486.1985.tb00078.x}.

\leavevmode\hypertarget{ref-Cao2010}{}%
Cao, Jian, Thomas Calderon, Akhilesh Chandra, and Li Wang. 2010.
``Analyzing late SEC filings for differential impacts of IS and
accounting issues.'' \emph{International Journal of Accounting
Information Systems} 11 (3): 189--207.
\url{https://doi.org/10.1016/j.accinf.2010.07.010}.

\leavevmode\hypertarget{ref-Cazier2016}{}%
Cazier, Richard A., and Ray J. Pfeiffer. 2016. ``Why are 10-K filings so
long?'' \emph{Accounting Horizons} 30 (1): 1--21.
\url{https://doi.org/10.2308/acch-51240}.

\leavevmode\hypertarget{ref-chen2013}{}%
chen. 2013. ``Opinion mining for relating multiword subjective
expressions and annual earnings in US financial statements.''
\emph{Journal of Information Science and Engineering}.
\url{http://nccur.lib.nccu.edu.tw/handle/140.119/66213}.

\leavevmode\hypertarget{ref-Chen2011}{}%
Chen, Chien Liang, Chao Lin Liu, Yuan Chen Chang, and Hsiang Ping Tsai.
2011. ``Mining opinion holders and opinion patterns in US financial
statements.'' In \emph{Proceedings - 2011 Conference on Technologies and
Applications of Artificial Intelligence, Taai 2011}, 62--68.
\url{https://doi.org/10.1109/TAAI.2011.19}.

\leavevmode\hypertarget{ref-Chen2018}{}%
Chen, Yu, Rhaad M. Rabbani, Aparna Gupta, and Mohammed J. Zaki. 2018.
``Comparative text analytics via topic modeling in banking.'' In
\emph{2017 Ieee Symposium Series on Computational Intelligence, Ssci
2017 - Proceedings}, 2018-Janua:1--8.
\url{https://doi.org/10.1109/SSCI.2017.8280945}.

\leavevmode\hypertarget{ref-Christensen2013}{}%
Christensen, Theodore E., William G. Heninger, and Earl K. Stice. 2013.
``Factors associated with price reactions and analysts' forecast
revisions around SEC filings.'' \emph{Research in Accounting Regulation}
25 (2): 133--48. \url{https://doi.org/10.1016/j.racreg.2013.08.003}.

\leavevmode\hypertarget{ref-Clarkson1999}{}%
Clarkson, Peter M., Jennifer L. Kao, and Gordon D. Richardson. 1999.
``Evidence That Management Discussion and Analysis (MD\&A) is a Part of
a Firm's Overall Disclosure Package.'' \emph{Contemporary Accounting
Research} 16 (1): 111--34.
\url{https://doi.org/10.1111/j.1911-3846.1999.tb00576.x}.

\leavevmode\hypertarget{ref-Cong2008}{}%
Cong, Yu, Alexander Kogan, and Miklos A. Vasarhelyi. 2007. ``Extraction
of Structure and Content from the Edgar Database: A Template‐Based
Approach.'' \emph{Journal of Emerging Technologies in Accounting} 4 (1):
69--86. \url{https://doi.org/10.2308/jeta.2007.4.1.69}.

\leavevmode\hypertarget{ref-Courtis1998}{}%
Courtis, John K. 1998. ``Annual report readability variability: Tests of
the obfuscation hypothesis.'' \emph{Accounting, Auditing \&
Accountability Journal} 11 (4): 459--72.
\url{https://doi.org/10.1108/09513579810231457}.

\leavevmode\hypertarget{ref-debreceny2011flex}{}%
Debreceny, Roger S, Stephanie M Farewell, Maciej Piechocki, Carsten
Felden, Andre Gräning, and Alessandro d'Eri. 2011. ``Flex or Break?
Extensions in Xbrl Disclosures to the Sec.'' \emph{Accounting Horizons}
25 (4): 631--57.

\leavevmode\hypertarget{ref-DeFranco2011}{}%
De Franco, Gus, M. H. Franco Wong, and Yibin Zhou. 2011. ``Accounting
adjustments and the valuation of financial statement note information in
10-K filings.'' \emph{Accounting Review} 86 (5): 1577--1604.
\url{https://doi.org/10.2308/accr-10094}.

\leavevmode\hypertarget{ref-Diaz2017}{}%
Diaz, Jamie, Kenneth Njoroge, and Philip B. Shane. 2017. ``Do Financial
Analysts Generate Value-Relevant Interpretive Information from 10-K
Filings?'' \url{https://doi.org/10.2139/ssrn.2967791}.

\leavevmode\hypertarget{ref-Dontoh2015}{}%
Dontoh, Alex, and Samir Trabelsi. 2015. ``Market Reaction to XBRL
Filings.'' \url{https://doi.org/10.2139/ssrn.2547579}.

\leavevmode\hypertarget{ref-Drake2012}{}%
Drake, Michael S., Darren T. Roulstone, and Jacob R. Thornock. 2012.
``What Investors Want: Evidence from Investors' Use of the EDGAR
Database.'' \emph{SSRN Electronic Journal}.
\url{https://doi.org/10.2139/ssrn.1932315}.

\leavevmode\hypertarget{ref-Drake2015}{}%
---------. 2015. ``The Determinants and Consequences of Information
Acquisition via EDGAR.'' \emph{Contemporary Accounting Research} 32 (3):
1128--61. \url{https://doi.org/10.1111/1911-3846.12119}.

\leavevmode\hypertarget{ref-DuarteSilva2013}{}%
Duarte‐Silva, Tiago, Huijing Fu, Christopher F. Noe, and K. Ramesh.
2013. ``How Do Investors Interpret Announcements of Earnings Delays?''
\url{https://doi.org/10.1111/j.1745-6622.2013.12007.x}.

\leavevmode\hypertarget{ref-Enev2017}{}%
Enev, Maria. 2017. ``Going Concern Opinions and Management's Forward
Looking Disclosures: Evidence from the MD\&A.''
\url{https://doi.org/10.2139/ssrn.2938703}.

\leavevmode\hypertarget{ref-engelberg2008costly}{}%
Engelberg, Joseph. 2008. ``Costly Information Processing: Evidence from
Earnings Announcements.'' In \emph{AFA 2009 San Francisco Meetings
Paper}.

\leavevmode\hypertarget{ref-Ertugrul2016}{}%
Ertugrul, Mine, Jin Lei, Jiaping Qiu, and Chi Wan. 2016. ``Annual Report
Readability, Tone Ambiguity, and the Cost of Borrowing.'' \emph{Ssrn} 52
(2): 811--36. \url{https://doi.org/10.2139/ssrn.2432797}.

\leavevmode\hypertarget{ref-Feldman2008}{}%
Feldman, Ronen, Suresh Govindaraj, Joshua Livnat, and Benjamin Segal.
2008. ``The Incremental Information Content of Tone Change in Management
Discussion and Analysis.'' \url{https://doi.org/10.2139/ssrn.1126962}.

\leavevmode\hypertarget{ref-Feldman2010}{}%
---------. 2010. ``Management's tone change, post earnings announcement
drift and accruals.'' \emph{Review of Accounting Studies} 15 (4):
915--53. \url{https://doi.org/10.1007/s11142-009-9111-x}.

\leavevmode\hypertarget{ref-Fengli2013}{}%
Fengli, Li, Russell Lundholm, and Minnis Michael. 2013. ``A measure of
competition based on 10-k filings.'' \emph{Journal of Accounting
Research} 51 (2): 399--436.
\url{https://doi.org/10.1111/j.1475-679X.2012.00472.x}.

\leavevmode\hypertarget{ref-fisher2010role}{}%
Fisher, Ingrid E, Margaret R Garnsey, Sunita Goel, and Kinsun Tam. 2010.
``The Role of Text Analytics and Information Retrieval in the Accounting
Domain.'' \emph{Journal of Emerging Technologies in Accounting} 7 (1):
1--24.

\leavevmode\hypertarget{ref-Foster2006}{}%
Foster, M. J., and David Hussey. 2006. ``The Truth about Corporate
Planning.'' \emph{The Journal of the Operational Research Society} 35
(4): 364. \url{https://doi.org/10.2307/2581178}.

\leavevmode\hypertarget{ref-Gandhi2017}{}%
Gandhi, Priyank, Tim Loughran, and Bill McDonald. 2017. ``Using Annual
Report Sentiment as a Proxy for Financial Distress in U.S. Banks.''
\emph{Ssrn}, March, 1--13. \url{https://doi.org/10.2139/ssrn.2905225}.

\leavevmode\hypertarget{ref-Ganguly2017}{}%
Ganguly, Arup. 2017. ``Textual Disclosure in SEC Filings and Litigation
Risk.''
\url{https://breakthroughanalysis.com/2008/08/01/unstructured-data-and-the-80-percent-rule/}.

\leavevmode\hypertarget{ref-Gibbons2018}{}%
Gibbons, Brian, Peter Iliev, and Jonathan Kalodimos. 2018. ``Analyst
Information Acquisition via EDGAR.'' \emph{Ssrn}, March.
\url{https://doi.org/10.2139/ssrn.3112761}.

\leavevmode\hypertarget{ref-Guo2017}{}%
Guo, Li, Feng Shi, and Jun Tu. 2017. ``Textual analysis and machine
leaning: Crack unstructured data in finance and accounting.'' \emph{The
Journal of Finance and Data Science} 2 (3): 153--70.
\url{https://doi.org/10.1016/j.jfds.2017.02.001}.

\leavevmode\hypertarget{ref-Han2016}{}%
Han, Meng, Yi Liang, Zhuojun Duan, and Yingjie Wang. 2016. ``Mining
public business knowledge: A case study in SEC's EDGAR.'' In
\emph{Proceedings - 2016 Ieee International Conferences on Big Data and
Cloud Computing, Bdcloud 2016, Social Computing and Networking,
Socialcom 2016 and Sustainable Computing and Communications, Sustaincom
2016}, 393--400.
\url{https://doi.org/10.1109/BDCloud-SocialCom-SustainCom.2016.65}.

\leavevmode\hypertarget{ref-Hajek2015}{}%
Hájek, Petr, and Vladimír Olej. 2015. ``Word categorization of corporate
annual reports for bankruptcy prediction by machine learning methods.''
In \emph{Lecture Notes in Computer Science (Including Subseries Lecture
Notes in Artificial Intelligence and Lecture Notes in Bioinformatics)},
9302:122--30. \url{https://doi.org/10.1007/978-3-319-24033-6_14}.

\leavevmode\hypertarget{ref-Heath1984}{}%
Heath, Robert L., and Greg Phelps. 1984. ``Annual reports II:
Readability of reports vs. business press.'' \emph{Public Relations
Review} 10 (2): 56--62.
\url{https://doi.org/10.1016/S0363-8111(84)80007-7}.

\leavevmode\hypertarget{ref-Hendricks2017}{}%
Hendricks, Bradley E., Mark H. Lang, and Kenneth J. Merkley. 2017.
``Through the Eyes of the Founder: CEO Characteristics and Firms'
Regulatory Filings.'' \url{https://doi.org/10.2139/ssrn.2962806}.

\leavevmode\hypertarget{ref-Henry2014}{}%
Henry, Elaine, and Andrew J. Leone. 2014. ``Measuring the Tone of
Accounting and Financial Narrative.'' In, 36--47.
\url{https://doi.org/10.4018/978-1-4666-4999-6.ch003}.

\leavevmode\hypertarget{ref-Henselmann2013}{}%
Henselmann, Klaus, Dominik Ditter, and Elisabeth Scherr. 2013.
``Irregularities in Accounting Numbers and Earnings Management -- A
Novel Approach Based on SEC XBRL Filings.''
\url{https://doi.org/10.2139/ssrn.2297355}.

\leavevmode\hypertarget{ref-hillegeist2004assessing}{}%
Hillegeist, Stephen A, Elizabeth K Keating, Donald P Cram, and Kyle G
Lundstedt. 2004. ``Assessing the Probability of Bankruptcy.''
\emph{Review of Accounting Studies} 9 (1): 5--34.

\leavevmode\hypertarget{ref-Hoitash2018}{}%
Hoitash, Rani, and Udi Hoitash. 2018. ``Measuring accounting reporting
complexity with XBRL.'' \emph{Accounting Review} 93 (1): 259--87.
\url{https://doi.org/10.2308/accr-51762}.

\leavevmode\hypertarget{ref-Holder2016}{}%
Holder, Anthony, Khondkar Karim, Karen (Jingrong) Lin, and Robert
Pinsker. 2016. ``Do material weaknesses in information
technology-related internal controls affect firms' 8-K filing timeliness
and compliance?'' \emph{International Journal of Accounting Information
Systems} 22: 26--43. \url{https://doi.org/10.1016/j.accinf.2016.07.003}.

\leavevmode\hypertarget{ref-Hsieh2013}{}%
Hsieh, Chia‐Chun, and Kai Wai Hui. 2013. ``Analyst Report Readability,
Earnings Uncertainty and Stock Returns.''
\url{https://doi.org/10.2139/ssrn.2182422}.

\leavevmode\hypertarget{ref-Hsieh2016}{}%
Hsieh, Chia Chun, Kai Wai Hui, and Yao Zhang. 2016. ``Analyst Report
Readability and Stock Returns.'' \emph{Journal of Business Finance and
Accounting} 43 (1-2): 98--130. \url{https://doi.org/10.1111/jbfa.12166}.

\leavevmode\hypertarget{ref-huizinga2012bank}{}%
Huizinga, Harry, and Luc Laeven. 2012. ``Bank Valuation and Accounting
Discretion During a Financial Crisis.'' \emph{Journal of Financial
Economics} 106 (3): 614--34.

\leavevmode\hypertarget{ref-Jackson2014}{}%
Jackson, Robert J., and Joshua Mitts. 2014. ``How the SEC Helps Speedy
Traders.'' \url{https://doi.org/10.2139/ssrn.2520105}.

\leavevmode\hypertarget{ref-Ji2016}{}%
Ji, Yuan, and Liang Tan. 2016. ``Labor Unemployment Concern and
Corporate Discretionary Disclosure.'' \emph{Journal of Business Finance
and Accounting} 43 (9-10): 1244--79.
\url{https://doi.org/10.1111/jbfa.12212}.

\leavevmode\hypertarget{ref-Khalil2017}{}%
Khalil, Samer, Sattar Mansi, Mohamad Mazboudi, and Andrew (Jianzhong)
Zhang. 2017. ``Bond Market Reaction to Untimely Filings of 10-K and 10-Q
Reports.'' \url{https://doi.org/10.2139/ssrn.3038837}.

\leavevmode\hypertarget{ref-Kim2012}{}%
Kim, Joung W., Jee-Hae Lim, and Won Gyun No. 2012. ``The Effect of First
Wave Mandatory XBRL Reporting across the Financial Information
Environment.'' \emph{Journal of Information Systems} 26 (1): 127--53.
\url{https://doi.org/10.2308/isys-10260}.

\leavevmode\hypertarget{ref-Kubick2016}{}%
Kubick, Thomas R., and G. Brandon Lockhart. 2016. ``Proximity to the SEC
and Stock Price Crash Risk.'' \emph{Financial Management} 45 (2):
341--67. \url{https://doi.org/10.1111/fima.12122}.

\leavevmode\hypertarget{ref-Lehavy2011}{}%
Lehavy, Reuven, Feng Li, and Kenneth Merkley. 2011. ``The effect of
annual report readability on analyst following and the properties of
their earnings forecasts.'' \emph{Accounting Review} 86 (3): 1087--1115.
\url{https://doi.org/10.2308/accr.00000043}.

\leavevmode\hypertarget{ref-Lerman2010}{}%
Lerman, Alina, and Joshua Livnat. 2010. ``The new Form 8-K
disclosures.'' \emph{Review of Accounting Studies} 15 (4): 752--78.
\url{https://doi.org/10.1007/s11142-009-9114-7}.

\leavevmode\hypertarget{ref-LeWitt1988}{}%
LeWitt, P. 1988. ``Hyperhidrosis and hypothermia responsive to
oxybutynin.'' \emph{Neurology} 38 (3): 506--7.
\url{http://www.ncbi.nlm.nih.gov/pubmed/3347362}.

\leavevmode\hypertarget{ref-Li2009}{}%
Li, Edward Xuejun, Ben Lansford, Joshua Livnat, Karen Nelson, Kathy
Petroni, Min Shen, Jake Thomas, and Jeff Wooldridge. 2009. ``Market
Reaction Surrounding the Filing of Periodic SEC Reports.'' \emph{Review
Literature and Arts of the Americas} 84 (4): 1171--1208.

\leavevmode\hypertarget{ref-Li2006}{}%
Li, Feng. 2006. ``Do Stock Market Investors Understand the Risk
Sentiment of Corporate Annual Reports?''
\url{https://doi.org/10.2139/ssrn.898181}.

\leavevmode\hypertarget{ref-Li2008}{}%
---------. 2008. ``The Determinants and Information Content of the
Forward-looking Statements in Corporate Filings - A Naive Bayesian
Machine Learning Approach.'' \url{https://doi.org/10.2139/ssrn.1267235}.

\leavevmode\hypertarget{ref-Li2010a}{}%
---------. 2010a. ``Managers' Self-Serving Attribution Bias and
Corporate Financial Policies.''
\url{https://doi.org/10.2139/ssrn.1639005}.

\leavevmode\hypertarget{ref-Li2010}{}%
---------. 2010b. ``The information content of forward- looking
statements in corporate filings-A naïve bayesian machine learning
approach.'' \emph{Journal of Accounting Research} 48 (5): 1049--1102.
\url{https://doi.org/10.1111/j.1475-679X.2010.00382.x}.

\leavevmode\hypertarget{ref-Li2014a}{}%
Li, Jun, and Xiaofei Zhao. 2014. ``Complexity and Information Content of
Financial Disclosures: Evidence from Evolution of Uncertainty Following
10-K Filings.'' \emph{SSRN Electronic Journal}.
\url{https://doi.org/10.2139/ssrn.2535469}.

\leavevmode\hypertarget{ref-Lim2018}{}%
Lim, Edwin Kia Yang, Keryn Chalmers, and Dean Hanlon. 2018. ``The
influence of business strategy on annual report readability.''
\emph{Journal of Accounting and Public Policy} 37 (1): 65--81.
\url{https://doi.org/10.1016/j.jaccpubpol.2018.01.003}.

\leavevmode\hypertarget{ref-Liu2018}{}%
Liu, Yu-Wen, Liang-Chih Liu, Chuan-Ju Wang, and Ming-Feng Tsai. 2018.
``RiskFinder: A Sentence-level Risk Detector for Financial Reports.''
In, 81--85. \url{https://doi.org/10.18653/v1/n18-5017}.

\leavevmode\hypertarget{ref-Lo2017}{}%
Lo, Kin, Felipe Ramos, and Rafael Rogo. 2017. ``Earnings management and
annual report readability.'' \emph{Journal of Accounting and Economics}
63 (1): 1--25. \url{https://doi.org/10.1016/j.jacceco.2016.09.002}.

\leavevmode\hypertarget{ref-Lopatta2017}{}%
Lopatta, Kerstin, Mario Albert Gloger, and Reemda Jaeschke. 2017. ``Can
Language Predict Bankruptcy? The Explanatory Power of Tone in 10-K
Filings.'' \emph{Accounting Perspectives} 16 (4): 315--43.
\url{https://doi.org/10.1111/1911-3838.12150}.

\leavevmode\hypertarget{ref-Lopatta2014}{}%
Lopatta, Kerstin, Reemda Jaeschke, and Cheong Yi. 2014. ``The Strategic
use of Language in Corrupt Firms' Financial Disclosures.'' \emph{Ssrn},
no. December. \url{https://doi.org/10.2139/ssrn.2512323}.

\leavevmode\hypertarget{ref-Loughran2016}{}%
Loughran, Tim, and Bill Mcdonald. 2016. ``Textual Analysis in Accounting
and Finance: A Survey.'' \emph{Journal of Accounting Research} 54 (4):
1187--1230. \url{https://doi.org/10.1111/1475-679X.12123}.

\leavevmode\hypertarget{ref-Loughran2017}{}%
Loughran, Tim, and Bill McDonald. 2017. ``The Use of EDGAR Filings by
Investors.'' \emph{Journal of Behavioral Finance} 18 (2): 231--48.
\url{https://doi.org/10.1080/15427560.2017.1308945}.

\leavevmode\hypertarget{ref-Loughran2009}{}%
---------. 2009. ``Plain English , Readability , and 10-K Filings.''
\emph{English}.

\leavevmode\hypertarget{ref-Loughran2014}{}%
---------. 2014. ``Regulation and financial disclosure: The impact of
plain English.'' \emph{Journal of Regulatory Economics} 45 (1): 94--113.
\url{https://doi.org/10.1007/s11149-013-9236-5}.

\leavevmode\hypertarget{ref-Loughran2015}{}%
---------. 2015. ``The Use of Word Lists in Textual Analysis.''
\emph{Journal of Behavioral Finance} 16 (1): 1--11.
\url{https://doi.org/10.1080/15427560.2015.1000335}.

\leavevmode\hypertarget{ref-Mangold2013}{}%
Mangold, Nancy R, Ching-lih Jan, John Tan, and Yi-pei Chen. 2013.
``Capital Market Effect of Mandatory XBRL Reporting: An Analysis of the
Phase-In Reporting Using Amended SEC Filings.'' \emph{International
Research Journal of Apllied Finance} IV (10): 1260--77.

\leavevmode\hypertarget{ref-Matthies2015}{}%
Matthies, Benjamin, and André Coners. 2015. ``Computer-aided text
analysis of corporate disclosures - Demonstration and evaluation of two
approaches.'' \emph{International Journal of Digital Accounting
Research} 15: 69--98. \url{https://doi.org/10.4192/1577-8517-v15_3}.

\leavevmode\hypertarget{ref-Mayew2015}{}%
Mayew, William J., Mani Sethuraman, and Mohan Venkatachalam. 2015.
``MD\&A disclosure and the firm's ability to continue as a going
concern.'' \emph{Accounting Review} 90 (4): 1621--51.
\url{https://doi.org/10.2308/accr-50983}.

\leavevmode\hypertarget{ref-Moffitt2009}{}%
Moffitt, Kevin, and Mary Burns. 2009. ``What Does That Mean?
Investigating Obfuscation and Readability Cues as Indicators of
Deception in Fraudulent Financial Reports.'' \emph{AMCIS 2009
Proceedings}.

\leavevmode\hypertarget{ref-MousaAlbashrawi2016}{}%
Mousa Albashrawi. 2016. ``Detecting Financial Fraud Using Data Mining
Techniques: A Decade Review from 2004 to 2015.'' \emph{Journal of Data
Science} 14 (3): 553--70. \url{http://www.jds-online.com/file}.

\leavevmode\hypertarget{ref-Qian2013}{}%
Qian, Buyue, and Hongfei Li. 2013. ``Does a company has bright future?
Predicting financial risk from revenue reports.'' In \emph{Proceedings
of 2013 Ieee International Conference on Service Operations and
Logistics, and Informatics}, 424--29. IEEE.
\url{https://doi.org/10.1109/SOLI.2013.6611452}.

\leavevmode\hypertarget{ref-Qiu2007}{}%
Qiu, Xin Ying. 2007. ``On building predictive models with company annual
reports.'' \emph{Phd Thesis, University of Iowa}.

\leavevmode\hypertarget{ref-rajan2015failure}{}%
Rajan, Uday, Amit Seru, and Vikrant Vig. 2015. ``The Failure of Models
That Predict Failure: Distance, Incentives, and Defaults.''
\emph{Journal of Financial Economics} 115 (2): 237--60.

\leavevmode\hypertarget{ref-rawte2018analysis}{}%
Rawte, Vipula, Aparna Gupta, and Mohammed J Zaki. 2018a. ``Analysis of
Year-over-Year Changes in Risk Factors Disclosure in 10-K Filings.'' In
\emph{Proceedings of the Fourth International Workshop on Data Science
for Macro-Modeling with Financial and Economic Datasets}, 1--4.

\leavevmode\hypertarget{ref-Rawte2018}{}%
Rawte, Vipula, Aparna Gupta, and Mohammed J. Zaki. 2018b. ``Analysis of
year-over-year changes in Risk Factors Disclosure in 10-K filings.'' In
\emph{Proceedings of the Fourth International Workshop on Data Science
for Macro-Modeling with Financial and Economic Datasets - Dsmm'18},
1--4. New York, New York, USA: ACM Press.
\url{https://doi.org/10.1145/3220547.3220555}.

\leavevmode\hypertarget{ref-Rogers2011}{}%
Rogers, Jonathan L., Andrew Van Buskirk, and Sarah L C Zechman. 2011.
``Disclosure tone and shareholder litigation.'' \emph{Accounting Review}
86 (6): 2155--83. \url{https://doi.org/10.2308/accr-10137}.

\leavevmode\hypertarget{ref-Ryans2017}{}%
Ryans, James. 2017. ``Using the EDGAR Log File Data Set.''
\url{https://doi.org/10.2139/ssrn.2913612}.

\leavevmode\hypertarget{ref-Sandulescu2015}{}%
Sandulescu, Paula Mirela. 2015. ``Insiders' incentives of using a
specific disclosure tone when trading.'' \emph{Studies in Communication
Sciences} 15 (1): 12--36.
\url{https://doi.org/10.1016/j.scoms.2015.03.009}.

\leavevmode\hypertarget{ref-Shirata2011}{}%
Shirata, Cindy Yoshiko, Hironori Takeuchi, Shiho Ogino, and Hideo
Watanabe. 2011. ``Extracting Key Phrases as Predictors of Corporate
Bankruptcy: Empirical Analysis of Annual Reports by Text Mining.''
\emph{Journal of Emerging Technologies in Accounting} 8 (1): 31--44.
\url{https://doi.org/10.2308/jeta-10182}.

\leavevmode\hypertarget{ref-Siano2018}{}%
Siano, Federico, and Peter D. Wysocki. 2018. ``The Primacy of Numbers in
Financial and Accounting Disclosures: Implications for Textual Analysis
Research.'' \emph{SSRN Electronic Journal}.
\url{https://doi.org/10.2139/ssrn.3223757}.

\leavevmode\hypertarget{ref-Smailovic2018}{}%
Smailović, Jasmina, Martin Žnidaršič, Aljoša Valentinčič, Igor
Lončarski, Marko Pahor, Pedro Tiago Martins, and Senja Pollak. 2018.
``Automatic Analysis of Annual Financial Reports: A Case Study.''
\emph{Computación Y Sistemas} 21 (4): 809--18.
\url{https://doi.org/10.13053/cys-21-4-2863}.

\leavevmode\hypertarget{ref-Smith2016}{}%
Smith, Kecia. 2016. ``Tell Me More: A Content Analysis of Expanded
Auditor Reporting in the United Kingdom.''
\url{https://doi.org/10.2139/ssrn.2821399}.

\leavevmode\hypertarget{ref-Song2017}{}%
Song, Shiwon. 2017. ``The Informational Value of Disaggregated Segment
Data: Evidence from the Textual Features of Business Descriptions.''
\emph{Ssrn}. \url{https://doi.org/10.2139/ssrn.3053791}.

\leavevmode\hypertarget{ref-Sourour2018}{}%
Sourour, Ben Saad, Msolli Badreddine, and Ajina Aymen. 2018. ``The
effect of annual report readability on financial analysts behaviour.''
\emph{Pressacademia} 5 (1): 26--37.
\url{https://doi.org/10.17261/pressacademia.2018.782}.

\leavevmode\hypertarget{ref-Stice1991}{}%
Stice, Earl K. 1991. ``The Market Reaction to 10-K and 10-Q filings and
to Subsequent The Wall Street Journal Earnings Accnouncements.''
\emph{The Accounting Review} 66 (1): 42--55.

\leavevmode\hypertarget{ref-Tao2018}{}%
Tao, Jie, Amit V. Deokar, and Ashutosh Deshmukh. 2018a. ``Analysing
forward-looking statements in initial public offering prospectuses: a
text analytics approach.'' \emph{Journal of Business Analytics} 1 (1):
54--70. \url{https://doi.org/10.1080/2573234x.2018.1507604}.

\leavevmode\hypertarget{ref-Tao2018analysing}{}%
Tao, Jie, Amit V Deokar, and Ashutosh Deshmukh. 2018b. ``Analysing
Forward-Looking Statements in Initial Public Offering Prospectuses: A
Text Analytics Approach.'' \emph{Journal of Business Analytics} 1 (1):
54--70.

\leavevmode\hypertarget{ref-tinoco2013financial}{}%
Tinoco, Mario Hernandez, and Nick Wilson. 2013. ``Financial Distress and
Bankruptcy Prediction Among Listed Companies Using Accounting, Market
and Macroeconomic Variables.'' \emph{International Review of Financial
Analysis} 30: 394--419.

\leavevmode\hypertarget{ref-Tsai2016}{}%
Tsai, Feng Tse, Hsin Min Lu, and Mao Wei Hung. 2016. ``The impact of
news articles and corporate disclosure on credit risk valuation.''
\emph{Journal of Banking and Finance} 68: 100--116.
\url{https://doi.org/10.1016/j.jbankfin.2016.03.018}.

\leavevmode\hypertarget{ref-Tsai2013}{}%
Tsai, Ming Feng, and Chuan Ju Wang. 2013. ``Risk ranking from financial
reports.'' In \emph{Lecture Notes in Computer Science (Including
Subseries Lecture Notes in Artificial Intelligence and Lecture Notes in
Bioinformatics)}, 7814 LNCS:804--7.
\url{https://doi.org/10.1007/978-3-642-36973-5_89}.

\leavevmode\hypertarget{ref-VanDenBogaerd2011}{}%
Van Den Bogaerd, MacHteld, and Walter Aerts. 2011. ``Applying machine
learning in accounting research.'' \emph{Expert Systems with
Applications} 38 (10): 13414--24.
\url{https://doi.org/10.1016/j.eswa.2011.04.172}.

\leavevmode\hypertarget{ref-wu2010comparison}{}%
Wu, Yanhui, Clive Gaunt, and Stephen Gray. 2010. ``A Comparison of
Alternative Bankruptcy Prediction Models.'' \emph{Journal of
Contemporary Accounting \& Economics} 6 (1): 34--45.

\leavevmode\hypertarget{ref-Xu2018}{}%
Xu, Qiao, Guy D. Fernando, and Kinsun Tam. 2018. ``Executive age and the
readability of financial reports.'' \emph{Advances in Accounting} 43:
70--81. \url{https://doi.org/10.1016/j.adiac.2018.09.004}.

\leavevmode\hypertarget{ref-Yang2018}{}%
Yang, Fang, Burak Dolar, and Lun Mo. 2018. ``Textual Analysis of
Corporate Annual Disclosures: A Comparison between Bankrupt and
Non-Bankrupt Companies.'' \emph{Journal of Emerging Technologies in
Accounting} 15 (1): 45--55. \url{https://doi.org/10.2308/jeta-52085}.

\leavevmode\hypertarget{ref-Yang2015}{}%
Yang, Shuo. 2015. ``The Disclosure and Valuation of Foreign Cash
Holdings.'' \url{https://doi.org/10.2139/ssrn.2558350}.

\leavevmode\hypertarget{ref-You2011}{}%
You, Haifeng, and Xiao Jun Zhang. 2011a. ``Limited attention and stock
price drift following earnings announcements and 10-K filings.''
\emph{China Finance Review International} 1 (4): 358--87.
\url{https://doi.org/10.1108/20441391111167487}.

\leavevmode\hypertarget{ref-You2011a}{}%
You, Haifeng, and Xiao-Jun Zhang. 2011b. ``Investor Under-Reaction to
Earnings Announcement and 10-K Report.'' \emph{SSRN Electronic Journal}.
\url{https://doi.org/10.2139/ssrn.1084332}.

\leavevmode\hypertarget{ref-You2009}{}%
You, Haifeng, and Xiao jun Zhang. 2009. ``Financial reporting complexity
and investor underreaction to 10-k information.'' \emph{Review of
Accounting Studies} 14 (4): 559--86.
\url{https://doi.org/10.1007/s11142-008-9083-2}.

\leavevmode\hypertarget{ref-Zhang2016}{}%
Zhang, Mei. 2016. ``Evaluation of Machine Learning Tools for
Distinguishing Fraud from Error.'' \emph{Journal of Business \&
Economics Research (JBER)} 11 (9): 393.
\url{https://doi.org/10.19030/jber.v11i9.8067}.
\end{cslreferences}

\bibliographystyle{unsrt}
\bibliography{edgar.bib}

\end{document}